%% file: main.tex
\documentclass{article}

\usepackage{microtype}
\usepackage{graphicx}
\usepackage{subcaption}
\usepackage{booktabs} %

\usepackage{hyperref}

\usepackage[preprint]{icml2026}

\usepackage{amsmath}
\usepackage{amssymb}
\usepackage{mathtools}
\usepackage{amsthm}
\usepackage[T1]{fontenc}

\usepackage[capitalize,noabbrev]{cleveref}

\theoremstyle{plain}

\theoremstyle{definition}

\theoremstyle{remark}

\usepackage[textsize=tiny]{todonotes}

\icmltitlerunning{\algoname: Joint Attribute Graphs for Filtered Nearest Neighbor Search}

\usepackage{enumitem}  %

\newcommand{\algoname}{JAG}

\usepackage{array}

\usepackage{graphicx}

\usepackage{float}
\usepackage{enumitem}

\usepackage{makecell}

\usepackage{booktabs}
\usepackage{pifont} %

\usepackage{subcaption}
\usepackage{makecell}

\usepackage{multirow}

\newcommand{\argmin}{\operatorname*{arg\,min}}

\newcommand{\cmark}{\textcolor{green!60!black}{\ding{51}}}%

\newcommand{\xmark}{\textcolor{red!70!black}{\ding{55}}}%
\newcommand{\myparagraph}[1]{\vspace{0.2em}\noindent{\bf \emph{#1.}} \,}

\AtBeginDocument{%
  \setlength{\abovedisplayskip}{4pt}
  \setlength{\belowdisplayskip}{4pt}
  \setlength{\abovedisplayshortskip}{2pt}
  \setlength{\belowdisplayshortskip}{2pt}
}

\begin{document}

\twocolumn[
  \icmltitle{\algoname: Joint Attribute Graphs for Filtered Nearest Neighbor Search}

  \icmlsetsymbol{equal}{*}

  \begin{icmlauthorlist}
    \icmlauthor{Haike Xu}{mit}
    \icmlauthor{Guy Blelloch}{cmu,google}
    \icmlauthor{Laxman Dhulipala}{umd,google}
    \icmlauthor{Lars Gottesbüren}{google}
    \icmlauthor{Rajesh Jayaram}{google}
    \icmlauthor{Jakub Łącki}{google}
  \end{icmlauthorlist}

  \icmlaffiliation{mit}{Massachusetts Institute of Technology}
  \icmlaffiliation{google}{Google Research}
  \icmlaffiliation{cmu}{Carnegie Mellon University}
  \icmlaffiliation{umd}{University of Maryland}

  \icmlcorrespondingauthor{Haike Xu}{haikexu@mit.edu}
  \icmlcorrespondingauthor{Rajesh Jayaram}{rkjayaram@google.com}

  \icmlkeywords{Machine Learning, ICML}

  \vskip 0.3in
]

\printAffiliationsAndNotice{}  %

\begin{abstract}
Despite filtered nearest neighbor search being a fundamental task in modern vector search systems, the performance of existing algorithms is highly sensitive to query selectivity and filter type. 
In particular, existing solutions excel either at specific filter categories (e.g., label equality) or within narrow selectivity bands (e.g., pre-filtering for low selectivity) and are therefore a poor fit for practical deployments that demand generalization to new filter types and unknown query selectivities.
In this paper, we propose \algoname~(Joint Attribute Graphs), a graph-based algorithm designed to deliver robust performance across the entire selectivity spectrum and support diverse filter types. 
Our key innovation is the introduction of \textit{attribute} and \textit{filter distances}, which transform binary filter constraints into continuous navigational guidance. 
By constructing a proximity graph that jointly optimizes for both vector similarity and attribute proximity, \algoname~prevents navigational dead-ends and allows \algoname{} to consistently outperform prior graph-based filtered nearest neighbor search methods.
Our experimental results across five datasets and four filter types (Label, Range, Subset, Boolean) demonstrate that \algoname{} significantly outperforms existing state-of-the-art baselines in both throughput and recall robustness\footnote{Our code is available at \url{https://github.com/xuhaike/JAG}}.
\end{abstract}

\vspace{-2em}

\section{Introduction}

\begin{figure}[!t]
    \centering
    \includegraphics[width=0.95\linewidth]{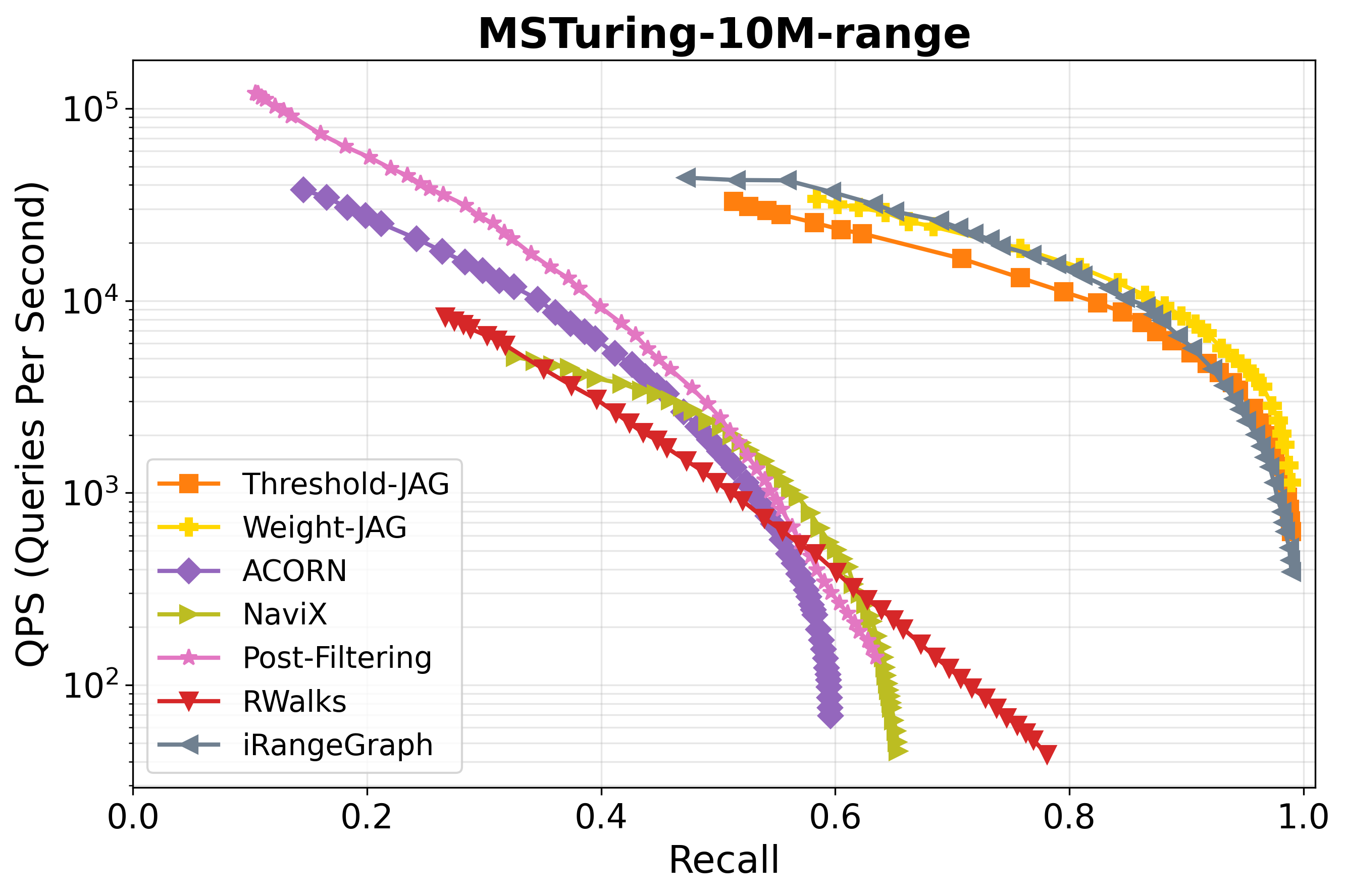}
    \caption{QPS vs. recall plot for range filters on the MSTuring-10M dataset. Please refer to Section~\ref{sec:experiments} for experimental details.}
    \label{fig:exp-msturing-range}
\end{figure}

Vector search has become a fundamental component of modern data management systems, driven by the increasing need to manage unstructured data and the wide popularity of deep learning embeddings.
While search over raw vector similarity is well-studied, practical vector search queries rarely rely on similarity alone.
Instead, users typically issue hybrid queries that combine high-dimensional vector similarity with structured metadata constraints. 
For example, a user might query a product catalog to find items visually similar to a reference image (which is a vector search query using the image embedding) but restrict the search to a specific price range or category.
This paradigm, broadly referred to as {\bf \emph{filtered vector search}}, introduces complex optimization trade-offs that differ significantly from standard relational query processing.

A critical parameter defining the performance landscape of filtered vector search is {\bf \emph{selectivity}}.
We explicitly define selectivity as the fraction of items in the index that satisfy the filter condition.
Consequently, a \textit{high selectivity} query implies that a large portion of the dataset satisfies the filter, whereas a \textit{low selectivity} query implies that only a small fraction of items are valid candidates. 
At the extreme ends of this selectivity spectrum, existing techniques provide robust solutions.
In the high selectivity regime (where most items are valid), {\bf \emph{post-filtering}} is generally effective.
This approach utilizes a standard, non-filtered vector search solution to retrieve a set of nearest neighbors, subsequently discarding those that fail the filter.
Since the filter is permissive, the probability of finding sufficient valid neighbors within the top-$k$ results remains high. 
Conversely, in the low selectivity regime (where very few items are valid), {\bf \emph{pre-filtering}} is the preferred strategy.
Here, the system identifies the valid subset of data first---a task for which modern databases are heavily optimized—and subsequently performs an exact search on the reduced candidate set.

However, a significant challenge arises in the wide selectivity range between these two extremes.
In this transition zone, the valid subset is too small for post-filtering to be efficient (requiring an excessively large scan of the vector index to find $k$ valid results) yet too large for pre-filtering to be performant (as the "reduced" set may still contain too many items for brute-force calculation).
Addressing this gap is an active area of study in the database community.
A variety of solutions have been proposed, ranging from specialized graph traversals~\cite{acorn, gollapudi2023filtered,ait2025rwalks,NHQ,li2025sieve} to hybrid indexing structures~\cite{gupta2023caps, mohoney2023high, zuo2024serf}.
These existing approaches often exhibit distinct performance profiles: some are optimized specifically for certain filter types (e.g., label equality~\cite{gollapudi2023filtered}, or range search~\cite{zuo2024serf,engels2024approximate}), while others excel primarily within specific sub-ranges of selectivity.
 
 To clarify the contributions of existing approaches and position our work, we classify hybrid search algorithms into three distinct categories based on their usage of metadata:
\begin{enumerate}[label=(\arabic*),topsep=0pt,itemsep=0pt,parsep=0pt,leftmargin=15pt]
    \item \textbf{Fully Oblivious (Filter Agnostic):} These methods (e.g., standard HNSW with post-filtering, ACORN~\cite{acorn}, NaviX~\cite{sehgal2025navix}) build the index solely on vector data, ignoring attributes during construction. They are universally applicable but suffer performance degradation when filters are restrictive, as the index does not guide the search toward valid items.
    
    \item \textbf{Filter-Aware:} These algorithms specialize the index structure for a specific class of query predicates known a priori, such as range queries~\cite{zuo2024serf,xu2024irangegraph} or label equality or disjunction based filters~\cite{gollapudi2023filtered,landrum2025ivf,cai2024navigating-ung}. 
    While they achieve good performance on their target query type, they are not robust to other types of queries: a range-optimized index cannot efficiently handle a Boolean or Subset query, requiring a different index for every filter type.
    
    \item \textbf{Attribute-Aware (also Filter Agnostic):} These methods~\cite{NHQ,ait2025rwalks} use the attribute data distribution to organize the index during construction but do not hard-code the structure for a specific filter logic. However, existing methods are still primarily designed for binary match/non-match filters (e.g., label equality) and are not sufficiently general to support arbitrary filter types. In contrast, our goal is to support diverse filter types—including but not limited to label, range, subset, and Boolean filters—within a single unified index structure.

\end{enumerate} 
 
Specialized \textit{filter-aware} algorithms will naturally outperform general-purpose methods on the specific task they are optimized for. 
However, modern workloads are dynamic and diverse and maintaining separate indices for every possible filter type is often computationally prohibitive.
Therefore, there is a strong incentive to design \textit{filter-agnostic} algorithms which are competitive across a wide range of filter types and selectivity ranges.

In this paper, we propose a new filtered vector search algorithm designed to operate robustly across the selectivity spectrum and support a diverse array of filter types. 
Our algorithm is thus \emph{filter-agnostic}, and is motivated by the observation that a given index configuration rarely suffices for all selectivity levels. 
Therefore, we introduce a method that effectively combines multiple index structures tuned for different selectivities into a single compact index.
In particular, our algorithm builds a {\bf \emph{Joint-Attribute Graph}}, or a {\bf \emph{\algoname}}, which is a \emph{single} graph-based index that integrates multiple navigational layers corresponding to different query selectivities.
To support this structural flexibility across broad filter types---including categorical, range, subset, and boolean filters---we formalize and utilize the concepts of \textit{attribute distance} and \textit{filter distance}.

The {\bf \emph{attribute distance}} ($dist_A$; used at build-time) measures the semantic proximity between the attributes of two data points (i.e., the metadata used to determine whether a datapoint satisfies the filter), independent of any specific query. 
By constructing the \algoname{} using a \emph{joint} metric of the vector distance (the distance between two vectors) and attribute distances, we ensure connectivity even in regions where the vector space is sparse but attributes are similar. 
Crucially, we employ a "capped" attribute distance mechanism when building a \algoname{}. 
By applying varying thresholds to $dist_A$, we generate a \emph{hierarchy} of navigational edges: some connecting strict attribute neighbors (essential for low selectivity) and others connecting broader attribute neighborhoods (sufficient for high selectivity). 
This unified index structure in a \algoname{} allows the search algorithm to seamlessly adapt its traversal strategy based on the selectivity of the query.

\begin{figure}[!t]
\centering
\includegraphics[width=0.9\linewidth]{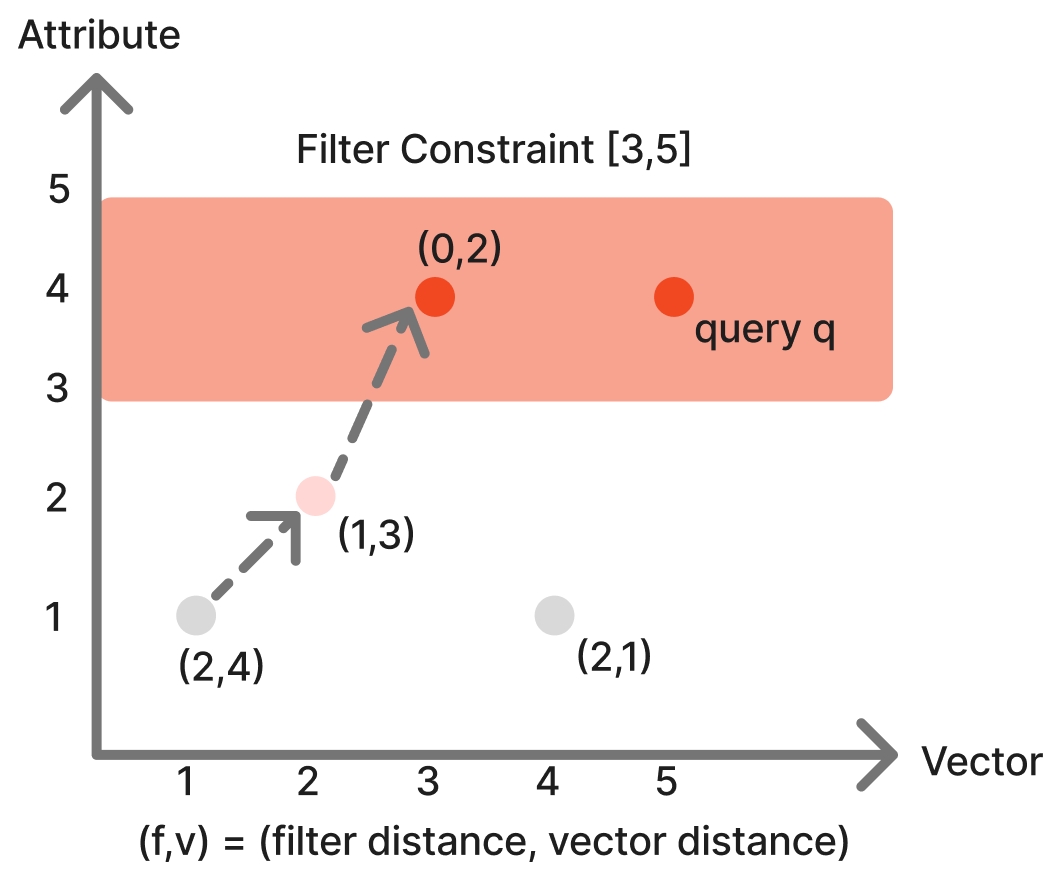}
\caption{An example illustrating how \algoname{} uses filter distance and vector distance to solve a range query. In this figure, both the vector value (x-axis) and attribute value (y-axis) are one-dimensional real numbers. The query specifies a filter range of [3,5]. The dashed arrow shows how filter distance and vector distance guide the greedy search toward the range query. Increasing intensity of the red color indicates improvement (decrease) in the filter distance.}
\label{fig:example}
\end{figure}

The {\bf \emph{filter distance}} ($dist_F$; used only at query-time when the filter is known) acts as a proxy for the binary filter constraint, quantifying how far a candidate's metadata is from satisfying the query.  
This prevents a query from hitting "dead ends" in the \algoname{} by providing navigational proxy even when neighbors do not strictly satisfy the filter. Please refer to an example from Figure~\ref{fig:example} to see how this proxy works.

Our experimental results using \algoname{} show that our ideas yield significant improvements over prior state-of-the-art filtered vector search algorithms.
For example, Figure~\ref{fig:exp-msturing-range} shows results on the MSTuring-10M dataset with range filters. All filter-agnostic algorithms reach at most 0.8 recall when their QPS is below 50, whereas our \algoname{} achieves QPS $>$ 10{,}000 at recall 0.8 and can still reach perfect recall with QPS $>$ 500. Furthermore, the performance of \algoname{} matches that of iRangeGraph, an algorithm specifically designed for range filters.

Our specific contributions are as follows:
\begin{enumerate}[label=(\arabic*),topsep=0pt,itemsep=0pt,parsep=0pt,leftmargin=15pt]
    \item We propose a unified graph-based indexing that maintains robust performance across the selectivity spectrum by integrating edges optimized for varying filter strictness.
    
    \item We formulate generalized attribute and filter distances that allow our index to support a wide range of constraints, including label, range, subset, and boolean logic, within a single framework.
    \item We demonstrate the superior performance of our approach through an extensive experimental evaluation across five datasets. Compared to state-of-the-art baselines, our method consistently achieves the best performance in terms of query throughput and recall among all filter-agnostic algorithms. Crucially, on challenging and standard datasets with mixed query workloads (e.g., LAION-25M and YFCC), \algoname{} achieves up to 10× higher QPS at recall 0.9 compared to the second-best filter-agnostic algorithm.
\end{enumerate}

\section{Preliminaries}

\myparagraph{Nearest Neighbors Search}
We define the \emph{nearest neighbor search} problem as follows. 
Let $X = \{x_1, \ldots, x_n\}$ be a set of $n$ points, where each $x \in \mathbb{R}^d$ is a $d$-dimensional vector. 
Given a query $x_q \in \mathbb{R}^d$, the goal of the nearest neighbor search is to design a data structure that can efficiently find
$\argmin_{x_p \in X} \operatorname{dist}(x_p,x_q)$,
where $\text{dist}(\cdot,\cdot)$ denotes a distance (or similarity) function. 
The quality of the returned candidate is typically evaluated by the \emph{approximation ratio} (from a theoretical perspective) or \emph{recall} (from an empirical perspective).

\myparagraph{Filtered Nearest Neighbors Search} 
In the filtered setting, each point $p = (x_p,a_p)$ comes with an attribute $a_p \in \mathcal{A}$, and each query $q = (x_q,f_q)$ comes with a filter $f_q \in \mathcal{F}$. 
The search must be performed under the constraint that the point’s attribute satisfies the filter, i.e.,
$g(a_p, f_q) = 1$,
where $g : \mathcal{A} \times \mathcal{F} \to \{0,1\}$ is a binary matching function. 
Formally, the goal is to find
$\argmin_{p \in P:\; g(a_p,f_q)=1}
    \operatorname{dist}(x_p,x_q)
$

\myparagraph{Filter Constraints}

Ideally, the filter constraint can be any function matching a pair of attribute and filter to pass / fail. We next specify several common forms of filter constraints considered in our work:

\begin{enumerate}[label=(\arabic*),topsep=0pt,itemsep=0pt,parsep=0pt,leftmargin=15pt]
    \item \textbf{Equality filter.}  
    $\mathcal{A} = [L]$, $\mathcal{F} = [L]$, and 
    \[
    g(a, f) = \mathbf{1}[a = f].
    \]
    This models categorical filtering, where each point belongs to exactly one category.

    \item \textbf{Range filter.}  
    $\mathcal{A} = \mathbb{R}$, $\mathcal{F} = \mathbb{R} \times \mathbb{R}$, and 
    \[
    g(a, f) = \mathbf{1}[f_{\min} \le a \le f_{\max}],
    \]
    where $f = (f_{\min}, f_{\max})$. 
    This captures numerical range filters.

    \item \textbf{Subset filter.}  
    $\mathcal{A} = \mathcal{F} = \{0,1\}^L$, and 
    \[
    g(a, f) = \mathbf{1}[f \wedge a = f],
    \]
    where $f \wedge a$ denotes elementwise AND. This models subset containment.

    \item \textbf{Boolean filter.}  
    $\mathcal{A} = \{0, 1\}^L$, $\mathcal{F}$ is a Boolean function over $L$ variables, and 
    \[
    g(a, f) = \mathbf{1}[f(a) = \texttt{1}].
    \]
    This represents general logical filters, where the filter is an arbitrary boolean predicate evaluated on the attributes.
\end{enumerate}

\section{Joint Attribute Graphs (JAG)}
In this section we detail the design and implementation of the Joint Attribute Graph (JAG), a unified indexing strategy capable of robust performance across varying selectivities. 
We first formalize the concepts of \emph{filter distance} and \emph{attribute distance}, which are essential for transforming binary filter constraints into continuous navigational gradients. 
We then explain how to integrate these metrics with standard vector distances using a capped attribute distance mechanism and a unified comparison rule to guide graph traversal. 
Finally, we present the specific algorithms for index construction and search, introducing our primary method, Threshold-JAG, alongside a weighted variant, Weight-JAG.

\subsection{Filter and Attribute Distances.}
Unlike the binary constraint function $g : \mathcal{A} \times \mathcal{F} \to \{0,1\}$ that only indicates whether an attribute--filter pair satisfies the constraint, 
we define two continuous distance functions:
\[
\text{(i)} \;
dist_F : \mathcal{A} \times \mathcal{F} \to \mathbb{R}_{\ge 0},
\;\text{and}\;
\text{(ii)} \;
dist_A : \mathcal{A} \times \mathcal{A} \to \mathbb{R}_{\ge 0},
\]
where $dist_F(f,a)$ measures how far an attribute $a$ is from satisfying a filter $f$, 
and $dist_A(a_1,a_2)$ measures how different two attributes are with respect to satisfying a (unknown at indexing time) filter.

\noindent
\textbf{Definition.}
The function $dist_F$ must satisfy the following two properties:
\begin{enumerate}[label=(\arabic*),topsep=0pt,itemsep=0pt,parsep=0pt,leftmargin=15pt]

    \item \textbf{Validity} For any attribute $a \in \mathcal{A}$ and filter $f \in \mathcal{F}$
    \[
    dist_F(a,f) = 0 \quad \Longleftrightarrow \quad g(a,f) = 1
    \]
    In other words, the filter distance is zero if and only if the attribute exactly satisfies the filter.
    \item \textbf{Consistency.}
    For any filter $f \in \mathcal{F}$ and attributes $a_1,a_2 \in \mathcal{A}$,
    \[
    dist_F(a_1,f) < dist_F(a_2,f)
    \]
    is consistent with the heuristic interpretation that $a_1$ is closer to satisfying $f$ than $a_2$.
    $dist_F(\cdot,f)$ induces an ordering over attributes that correlates with proximity to filter satisfaction, even though satisfaction is discrete.
\end{enumerate}

Similarly, the \emph{attribute distance} $dist_A$ must satisfy two analogous criteria:
\begin{enumerate}[label=(\arabic*),topsep=0pt,itemsep=0pt,parsep=0pt,leftmargin=15pt]
    \item \textbf{Validity.} $dist_A(a_1,a_2) = 0$ if and only if $a_1 = a_2$.
    \item \textbf{Consistency.}
    For any attributes $a_1,a_2,a_3 \in \mathcal{A}$,
    \[
    dist_A(a_1,a_2) < dist_A(a_1,a_3)
    \]
    is consistent with the heuristic interpretation that $a_1$ and $a_2$ are more likely to simultaneously pass or fail an unknown filter\footnotemark than $a_1$ and $a_3$. $dist_A$ induces an ordering over attribute pairs that correlates with similarity in filter outcomes.
\end{enumerate}

\footnotetext{This notion is somewhat informal, as the exact likelihood cannot be rigorously defined without knowing the filter in advance.}

These measures provide fine-grained relational information among attributes and filters, which can be exploited both when constructing proximity graphs over attributes and when routing queries through the graph.

\myparagraph{Examples}
The proposed distances can be naturally instantiated for several common filtering scenarios:
\begin{enumerate}[label=(\arabic*),topsep=0pt,itemsep=0pt,parsep=0pt,leftmargin=15pt]
 
    \item \textbf{Equality filter.} Attribute / filter represents a discrete category. Let $\mathcal{A}=\mathcal{F}=[L]$,
    We define
    {\small
    \[
    dist_F(a,f)=
    \begin{cases}
    0, & a=f,\\
    1, & a\neq f,
    \end{cases}
    \quad
    dist_A(a_1,a_2)=\mathbf{1}[a_1\neq a_2].
    \]
    }
    \item \textbf{Range filter.} Attributes are scalar values (e.g., timestamps, prices).
    Let $\mathcal{A}=\mathbb{R}$ and $\mathcal{F}=\mathbb{R}\times\mathbb{R}$ with a filter
    $f=(f_{\min},f_{\max})$.
    We define
    {\small
    \[
    \begin{aligned}
    dist_F(a,f) &=
    \begin{cases}
    f_{\min}-a, & a<f_{\min},\\
    0, & a\in[f_{\min},f_{\max}],\\
    a-f_{\max}, & a>f_{\max},
    \end{cases} \\
    \end{aligned}
    \]
    }
    \[
    dist_A(a_1,a_2) = |a_1-a_2|,
    \]
    
    where $dist_A$ measures numeric proximity between attributes. 

    \item \textbf{Subset filter.} Attributes encode set membership
    (e.g., tags, privileges). Let $\mathcal{A}=\mathcal{F}=\{0,1\}^L$.
    We define
    \[
    dist_F(a,f) = |f \setminus a|,
    \qquad
    dist_A(a_1,a_2) = |a_1 \oplus a_2|,
    \]
    where $\oplus$ denotes bitwise XOR. Here, $dist_F$ measures the number of elements $a$ needs to cover $f$, while $dist_A$ measures the number of elements they differ.

    \item \textbf{Boolean filter.} Attributes encode Boolean assignments over variables.
    Let $\mathcal{A}=\{0,1\}^L$, and let $\mathcal{F}$ denote the set of all Boolean predicates.
    We define
    \[
    \begin{aligned}
    dist_F(a,f) &= \min_{a' : f(a') = 1} \lvert a - a' \rvert, \\
    dist_A(a_1,a_2) &= \lvert a_1 - a_2 \rvert,
    \end{aligned}
    \]
    where $\lvert \cdot \rvert$ denotes Hamming distance. Here, $dist_F(a,f)$ measures the minimum number of bit flips required to satisfy the Boolean predicate $f$.
\end{enumerate}

\myparagraph{Discussion}
The specific choices of $dist_F$ and $dist_A$ are not unique and may vary across applications. 
In the worst case, one can always define trivial distances:
\[
dist_F(a,f)=\mathbf{1}[g(f,a)=0], \quad
dist_A(a_1,a_2)=\mathbf{1}[a_1\ne a_2]
\]
which guarantees the feasibility of incorporating filter and attribute distances under any filtering scheme.
Our algorithm is compatible with all such choices of $dist_F$ and $dist_A$, but the performance may vary depending on how informative the attribute/filter distance is.

\subsection{Merging Attribute and Standard Distances}
\label{sec:capped_dist}
Given the definitions of attribute distance $dist_A$ and filter distance $dist_F$, a natural idea is to construct a proximity graph over attributes using $dist_A$, and then perform a greedy search guided by $dist_F$. 
Intuitively, this should allow us to quickly locate points whose attributes are most likely to satisfy the query filter. 
However, two major challenges remain:

\begin{enumerate}[label=(\arabic*),topsep=0pt,itemsep=0pt,parsep=0pt,leftmargin=15pt]
    \item \textbf{Inconsistency between attribute similarity and filter constraints.}
    We define attribute similarity as a proxy for approximating the likelihood that two attributes will simultaneously pass an unknown filter. However, this approximation may be misleading if the index relies too heavily on attribute similarity. In the extreme case, an unfiltered query does not depend on attribute information at all. It is inherently difficult to approximate all unknown filters using a fixed attribute similarity. 

    \item \textbf{Combining attribute and standard proximity.}  
    It is nontrivial to integrate the attribute/filter proximity graph with the standard proximity graph over the data vectors. 
    Points that are close in the embedding space may have completely different attributes, 
    and vice versa, making it unclear how to unify the two distance notions.
\end{enumerate}

Prior work such as \cite{NHQ} and \cite{ait2025rwalks} has proposed constructing a proximity graph based on a 
weighted combination of the standard vector distance and the filter distance.
While intuitive, this approach faces two limitations:
(i)~the two distances are often incomparable in scale, as one typically arises from the Euclidean space (e.g., $\ell_2$ distance) 
while the other reflects discrete or categorical mismatches (e.g., keyword overlap); and 
(ii)~the optimal weighting may depend on the query selectivity, which is unknown at construction time.

\myparagraph{Our Solution}
To address these challenges, we propose a new approach that introduces the concept of \emph{capped attribute distance} 
and a unified comparison rule for candidate evaluation. 

\myparagraph{Capped attribute distance}  
We define a threshold-dependent variant of the attribute distance:
\[
dist_A(a_1, a_2; t) = \max(dist_A(a_1, a_2) - t, 0),
\]
where $t \ge 0$ is a tunable threshold. 
This formulation implies that once two attributes are sufficiently close (within distance $t$), 
they are treated as equivalent for indexing purposes. 
Conceptually, this introduces an ``indicator-like'' property: 
all attributes within the $t$-neighborhood are assumed to satisfy the "hypothetic" filter condition, 
while attributes farther away are penalized proportionally to their excess attribute distance.

\myparagraph{Unified comparison rule}
Given a base point $p = (x_p, a_p)$ and threshold $t$, we compare two candidate points $u = (x_u,a_u)$ and $v = (x_v,a_v)$ 
relative to $p$ by first evaluating their capped attribute distances $dist_A(a_{p}, a_{u}; t)$ 
and $dist_A(a_{p}, a_{v}; t)$. 
If the two candidates are equally close in attribute distance, 
we break ties using their standard vector distances $dist(x_p, x_u)$ and $dist(x_p, x_v)$.

At query time, given a query $q$ and two candidate points $u$ and $v$, 
we analogously compare them using $dist_F(f_q, a_u)$ and $dist_F(f_q, a_v)$, 
and if equal, break ties using their standard vector distances to $x_q$.

This design unifies discrete filter constraints and continuous vector distances under a single graph-based framework, allowing the search to adapt smoothly across different levels of query selectivity.

Formally, for any threshold parameter $t \ge 0$ and two points $u$ and $v$, we define the unified distance as an ordered pair
\[
D_A^t(u, v)
    = \bigl( dist_A(a_u, a_v; t), \, dist(x_u, x_v) \bigr),
\]
Similarly, for a query $q$ and a point $u$, we define
\[
D_F(q,u)
    = \bigl( dist_F(f_q, a_u), \, dist(x_q, x_u) \bigr),
\]
We compare $D_A^t(u,v)$ or $D_F(q,u)$ by lexicographic order.

\subsection{Threshold-\algoname}

To support filters with varying selectivities, we construct proximity graphs using a set of thresholds $T$ and perform greedy search over the resulting unified index. We refer to this version of our algorithm as Threshold-\algoname. 

Our index construction and query procedures rely on three components:
\textsf{GreedySearch}, \textsf{Insert}, and \textsf{JointRobustPrune}. 
All procedures operate on a directed proximity graph $G = (P, E)$ 
and make use of customized comparators that incorporate both attribute 
distance and filter distance as we have discussed in Subsection~\ref{sec:capped_dist}.

\textsf{{\bf GreedySearch}} (Algorithm~\ref{alg:search}):  
To find the nearest neighbors of a point $x$ (which can be either a query or an existing data point)
under a comparator $D$, the algorithm performs a greedy beam search on the graph starting from a fixed entry point $s$.
At each iteration, it expands the closest unexplored vertex $p$, 
computes the comparator
distances from $x$ to all out-neighbors $N_{out}(p)$ of $p$, 
and adds these neighbors to a candidate list maintained as a priority queue ordered by $D(x,\cdot)$. 
The search terminates when all the top-$l_s$ vertices in the queue have been explored.
Finally, it returns the $k$ closest explored vertices as the search result.

\textsf{{\bf Query}} (Algorithm~\ref{alg:query}):  
To answer a filtered nearest-neighbor query $(q,f)$, 
we invoke \textsf{GreedySearch} using a comparator $D_F(f)$ 
that prioritizes candidates according to their filter distances $dist_F$ (breaking ties by vector distance). 

\textsf{{\bf Insert}} (Algorithm~\ref{alg:insert}):  
The graph is built incrementally. 
To insert a new point $p$ with attribute $a_p$, 
we perform \textsf{GreedySearch} with comparator $D_A(t)$ for each threshold $t$ in a predefined list $T$, 
and collect the union of all visited vertices $V$. 
The candidate set $V$ is then pruned to a subset $V'$, $|V'|\le R$, 
by invoking \textsf{JointRobustPrune}. 
Bidirectional edges are established between $p$ and each $v\in V'$. 
If the degree of any vertex $v$ exceeds $R$, we apply \textsf{JointRobustPrune} again to enforce the degree constraint.

\textsf{{\bf JointRobustPrune}} (Algorithm~\ref{alg:pruning}):  
When a vertex exceeds the maximum out-degree $R$, 
this procedure selects a diverse subset of neighbors. 
Let $T$ be the list of thresholds maintained in the index and $\deg$ the total degree budget. 
We partition the degree budget into $|T|$ buckets, assigning $\deg/|T|$ neighbors per threshold.  
For each $t\in T$, we sort candidates by comparator $D_A(t)$, 
then iterate through the sorted list in order:  
if a candidate $v$ is not dominated by previously selected vertices, we include it in $V'$ and prune later vertices $v'$ 
for which $dist_V(v,v') < dist_V(p,v')/\alpha$ according to the standard RobustPruning criteria in \cite{DiskANN19}.  
The process continues until each bucket reaches its local degree limit $\deg/|T|$.  
Finally, we merge all buckets to form the new neighbor list $V'$, ensuring $|V'|\le \deg$.

\begin{algorithm}[t]
\caption{GreedySearch($G, q, k, l_s, D$)}
\label{alg:search}
\begin{algorithmic}[1]
\STATE \textbf{Input:} Graph $G=(P,E)$, query point $q$, beam size $l_s$, comparator $D$
\STATE \textbf{Output:} Top-$k$ nearest vertices to $q$, visited set $V$
\STATE Initialize candidate list $L \leftarrow \{s\}$ \COMMENT{$s$ is the entry vertex}
\STATE Initialize visited set $V \leftarrow \emptyset$
\WHILE{$L \setminus V \neq \emptyset$}
    \STATE $p \leftarrow \arg\min_{v \in L \setminus V} D(q,v)$
    \STATE $V \leftarrow V \cup \{p\}$
    \FOR{each $u \in N_{\text{out}}(p)$}
        \IF{$u \notin L$}
            \STATE $L \leftarrow L \cup \{u\}$
        \ENDIF
    \ENDFOR
    \IF{$|L| > l_s$}
        \STATE Retain the top-$l_s$ vertices in $L$ ranked by $D(q,\cdot)$
    \ENDIF
\ENDWHILE
\STATE \textbf{return} Top-$k$ vertices in $V$ ranked by $D(q,\cdot)$, and $V$
\end{algorithmic}
\end{algorithm}

\begin{algorithm}[t]
\caption{Query($G, q, l_s, k$)}
\label{alg:query}
\begin{algorithmic}[1]
\STATE \textbf{Input:} Graph $G=(P,E)$, query point $q$, search beam size $l_s$, number of returned neighbors $k$
\STATE \textbf{Output:} Top-$k$ nearest vertices to $q$
\STATE $[A, V] \leftarrow \textsf(\textsf{GreedySearch}(G, q, k, l_s, D_F))$
\STATE \textbf{return} $A$
\end{algorithmic}
\end{algorithm}

\begin{algorithm}[h]
\caption{Insert($G, p, T, l_b, R, \alpha$)}
\label{alg:insert}
\begin{algorithmic}[1]
\STATE \textbf{Input:} Graph $G=(P,E)$, new point $p$, threshold list $T$, build beam size $l_b$, degree bound $R$, pruning parameter $\alpha$
\STATE \textbf{Output:} Updated graph $G'=(P',E')$
\STATE Initialize $V \leftarrow \emptyset$
\FOR{each $t \in T$}
    \STATE $[A_t, V_t] \leftarrow \textsf{GreedySearch}(G, p, 1, l_b, D_A(t))$
    \STATE $V \leftarrow V \cup V_t$
\ENDFOR
\STATE $N_{\text{out}}(p) \leftarrow \textsf{JointRobustPrune}(G, p, V, R, \alpha, T)$
\FOR{each $v \in N_{\text{out}}(p)$}
    \STATE $N_{\text{out}}(v) \leftarrow N_{\text{out}}(v) \cup \{p\}$
    \IF{$|N_{\text{out}}(v)| > R$}
        \STATE $N_{\text{out}}(v) \leftarrow \textsf{JointRobustPrune}(G, v, N_{\text{out}}(v), R, \alpha, T)$
    \ENDIF
\ENDFOR
\STATE \textbf{return} $G'$
\end{algorithmic}
\end{algorithm}

\begin{algorithm}[h]
\caption{JointRobustPrune($G, p, V, R, \alpha, T$)}
\label{alg:pruning}
\begin{algorithmic}[1]
\STATE \textbf{Input:} Graph $G=(P,E)$, vertex $p$, candidate set $V$, total degree budget $R$, pruning parameter $\alpha$, threshold list $T$
\STATE \textbf{Output:} Pruned neighbor list $V'$
\STATE Initialize $V' \leftarrow \emptyset$
\FOR{each $t \in T$}
    \STATE Initialize $V_t' \leftarrow \emptyset$
    \STATE Sort $V$ in increasing order of $D_A(t)$
    \FOR{each $v \in V$ (in order)}
        \IF{$\forall u \in V_t',\ \alpha \cdot \mathrm{dist}(u,v) > \mathrm{dist}(p,v)$}
            \STATE $V_t' \leftarrow V_t' \cup \{v\}$
        \ENDIF
        \IF{$|V_t'| \ge R / |T|$}
            \STATE \textbf{break}
        \ENDIF
    \ENDFOR
    \STATE $V' \leftarrow V' \cup V_t'$
\ENDFOR
\STATE \textbf{return} $V'$
\end{algorithmic}
\end{algorithm}

\subsection{Weight-\algoname}
We also implement a weighted variant of \algoname, in which attribute 
distance and vector distance are combined using different weights. 
For a weight $w$, we define
\[
D_A^w(u,v) = w \cdot \mathrm{dist}_A(a_u, a_v) + \mathrm{dist}(x_u, x_v).
\]
We build the index graph using a set of such weights. Similar to Threshold-\algoname, when calling \textsf{Insert} and \textsf{JointRobustPrune}, we iterate over the weight list and use $D^w_A(u,v)$ as the comparator. During query processing, we still compare $\mathrm{dist}_F(\cdot)$ first, followed by the standard vector  distance. We refer to this variant as Weight-\algoname. Please refer to Appendix~\ref{sec:ablation} for experimental comparisons.

\section{Experiments}
\label{sec:experiments}

We compare our algorithm, \algoname, against ten baseline algorithms on five datasets. Please see our code in the supplementary material. We summarize our findings below.
\begin{itemize}[leftmargin=*, itemsep=0pt, topsep=0pt]
\item Across all datasets (SIFT, ARXIV, LAION, YFCC, and MSTuring), \algoname{} consistently outperforms all filter-agnostic algorithms (see Figure~\ref{fig:exp-msturing-range}, \ref{fig:exp-label}, \ref{fig:exp-range}, \ref{fig:exp-subset}).
\item For datasets with low-selectivity queries (e.g., selectivity $< 1/100$ on MSTuring), \algoname{} is the only filter-agnostic method that achieves perfect recall. All other filter-agnostic algorithms plateau below 0.8 recall, even when operating at extremely low throughput (QPS $< 50$), whereas \algoname{} attains QPS~$>1000$ at 0.8 recall (see Figure~\ref{fig:exp-msturing-range}, \ref{fig:exp-range}, \ref{fig:exp-subset}).
\item \algoname{} exhibits the best QPS performance as the data scale increases (see Figure~\ref{fig:exp-scaling}).
\item \algoname{} achieves the highest QPS when there is correlation between the query filter and the vector space (see Figure~\ref{fig:exp-correlation}).
\end{itemize}

\subsection{Datasets}

We evaluate our algorithms on five datasets under various filter setups, including~\textbf{SIFT}, \textbf{ARXIV}, \textbf{LAION}, \textbf{YFCC}, and \textbf{MSTuring}, each paired with one or more of the \emph{Label}, \emph{Range}, \emph{Subset}, and \emph{Boolean} filter types. We summarize the dataset size, average selectivity in Table~\ref{tab:exp-preprocessing}. Please refer to Appendix~\ref{sec:full-dataset} for details on all datasets.

\subsection{Algorithms}\label{sec:exp-alg}

We implement both Threshold-\algoname{} and Weight-\algoname{} in our experiment. Please refer to Section~\ref{sec:implementation_details} for implementation details. We compare our algorithm against the following baseline algorithms.

\myparagraph{Baselines}
We compare \algoname~with a comprehensive set of baseline algorithms, including
ACORN~\citep{acorn}, NaviX~\citep{sehgal2025navix}, RWalks~\citep{ait2025rwalks}, Post-Filtering. We also test some other baseline algorithms, which only support certain filters. For example, we test FilteredVamana~\citep{gollapudi2023filtered}, StitchedVamana~\citep{gollapudi2023filtered}, and UNG~\citep{cai2024navigating-ung} on Label and Subset filters, NHQ~\citep{NHQ} on Label filters, and iRangeGraph~\cite{xu2024irangegraph} on Range Filters. We summarize their compatibility in
Table~\ref{tab:alg-filter}. We describe each baseline algorithm in Appendix~\ref{sec:full-exp-alg} and their parameter choices in Appendix~\ref{sec:parameters}.

\begin{figure}[t]
    \centering
    \begin{subfigure}[t]{0.45\textwidth}
        \centering
        \includegraphics[width=\linewidth]{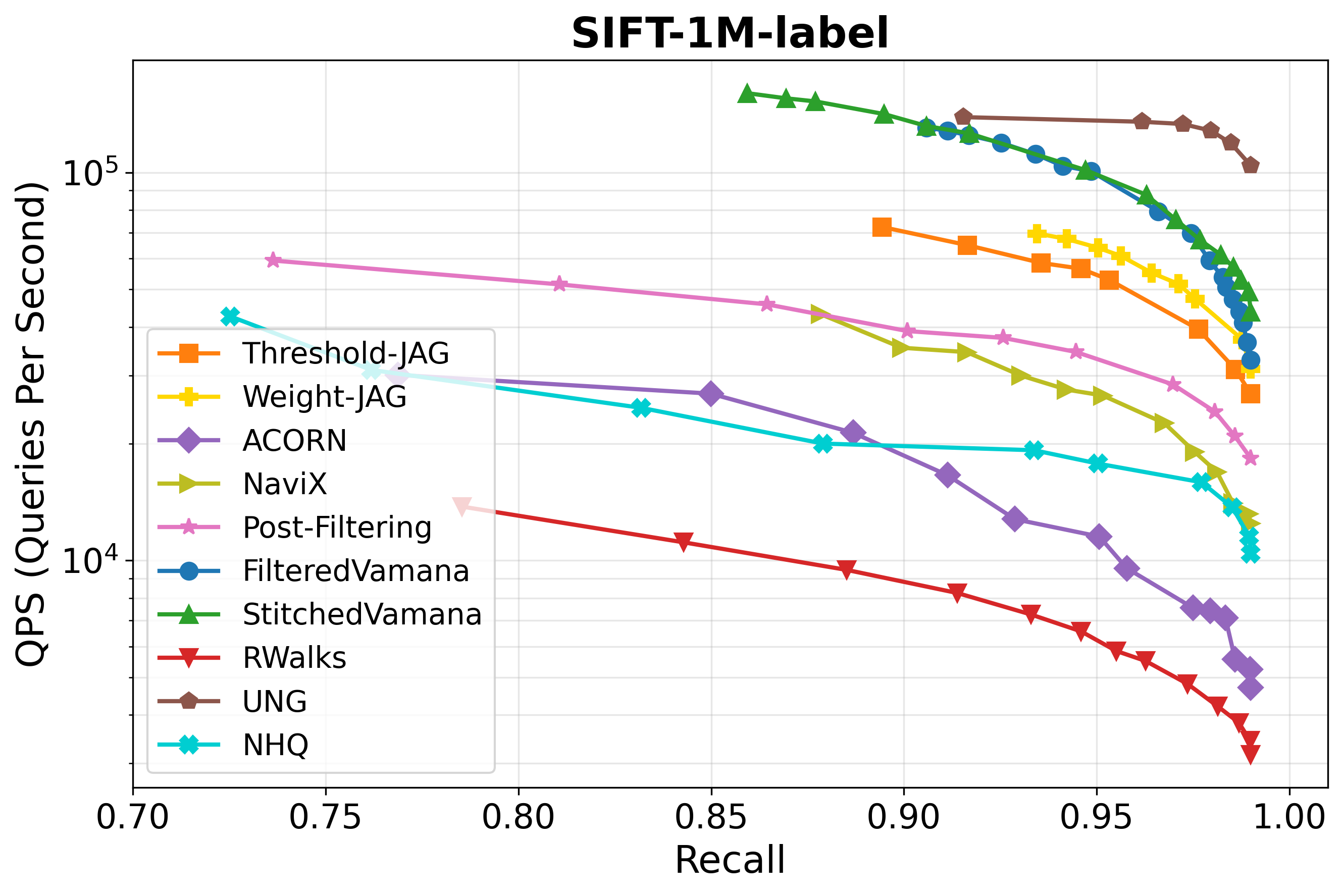}
        \label{fig:sift-label}
    \end{subfigure}
    \vspace{-1em}
    \\ 
    \begin{subfigure}[t]{0.45\textwidth}
        \centering
        \includegraphics[width=\linewidth]{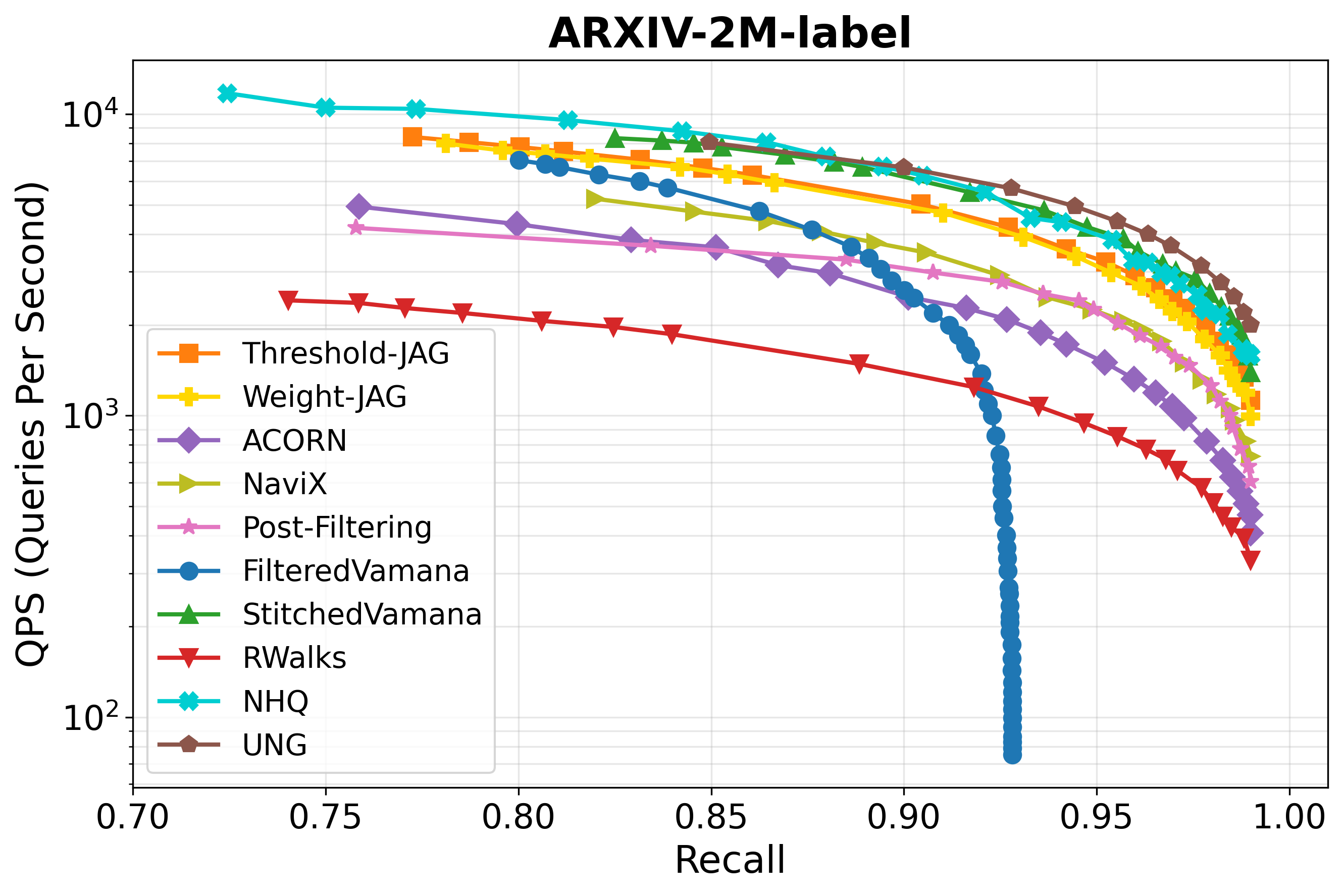}
        \label{fig:arxiv-label}
    \end{subfigure}
    \vspace{-1em}
    \caption{QPS vs. recall plot for Label filters on the SIFT and ARXIV datasets. Note that NHQ is designed specifically for Label filter. FilteredVamana, StitchedVamana, and UNG are designed specifically for Label and Subset filters.}
    \label{fig:exp-label}
\end{figure}

\begin{figure*}[t]

    \vspace{-0.5em}

    \centering
    
    \begin{subfigure}[t]{0.45\textwidth}
        \centering
        \includegraphics[width=\linewidth]{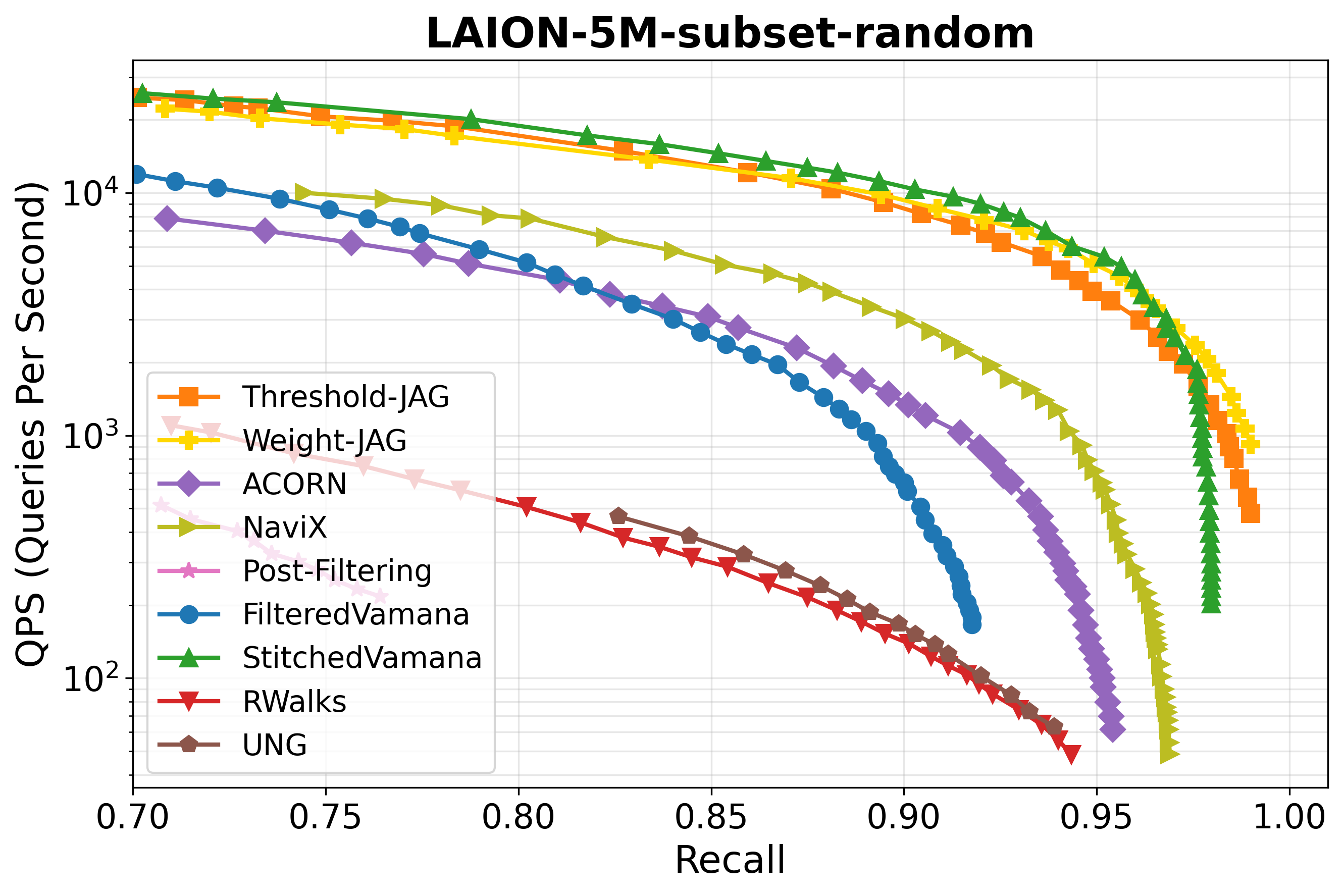}
        \label{fig:laion5m-subset}
    \end{subfigure}
    \hfill
    \begin{subfigure}[t]{0.45\textwidth}
        \centering
        \includegraphics[width=\linewidth]{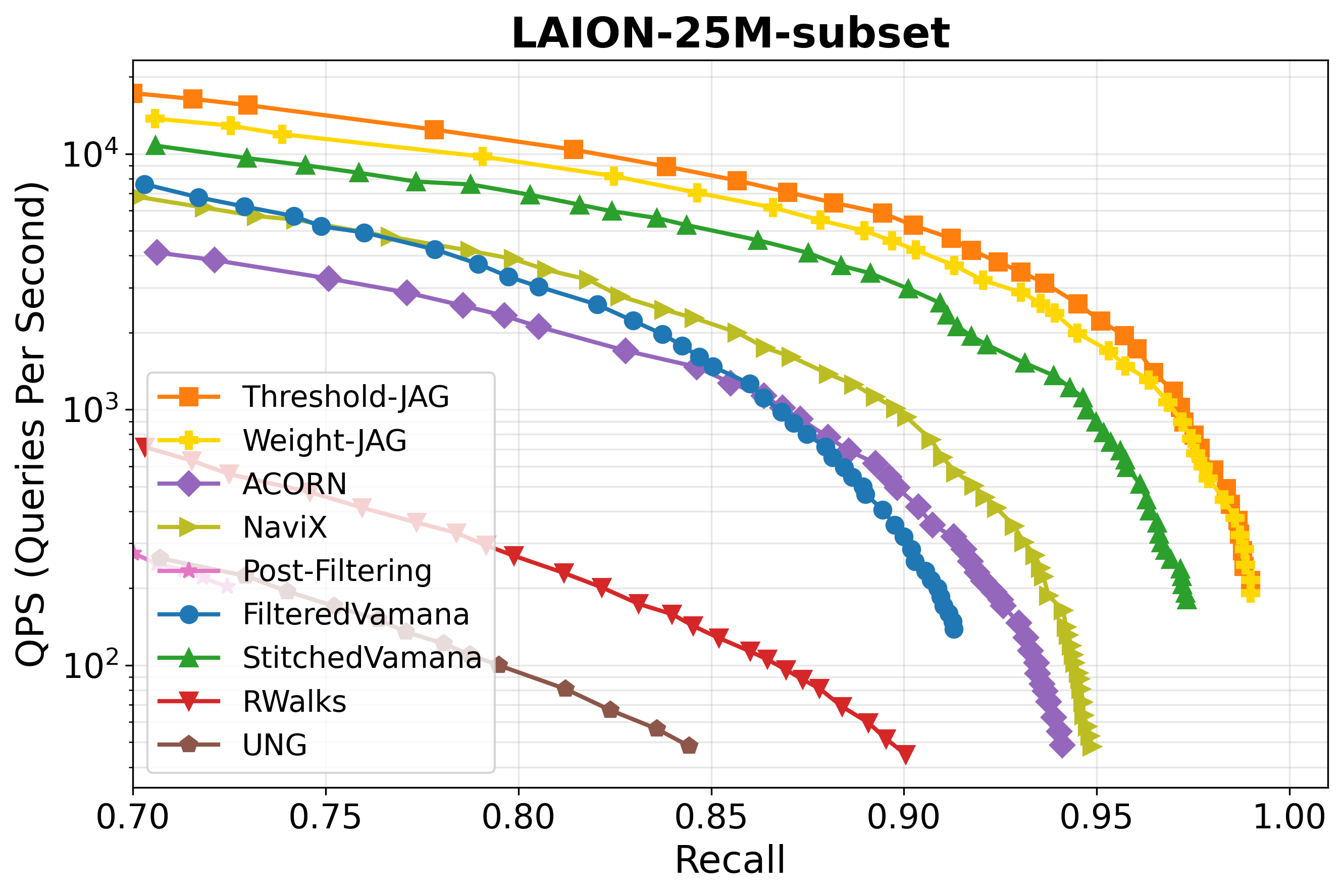}
        \label{fig:laion25m-subset}
    \end{subfigure}

    \vspace{-1.3em}

    \begin{subfigure}[t]{0.45\textwidth}
        \centering
        \includegraphics[width=\linewidth]{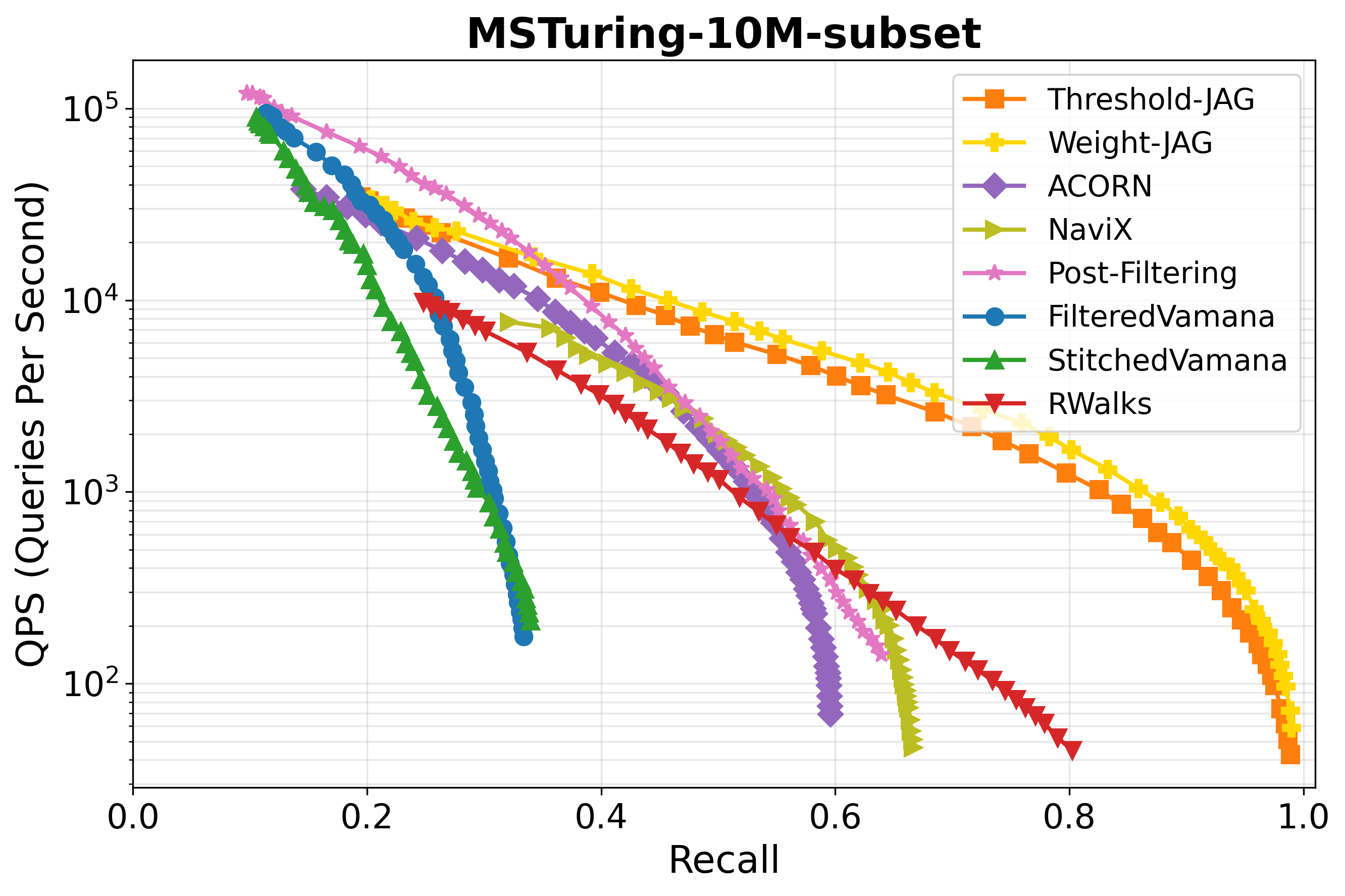}
        \label{fig:msturing-subset}
    \end{subfigure}
    \hfill
    \begin{subfigure}[t]{0.45\textwidth}
        \centering
        \includegraphics[width=\linewidth]{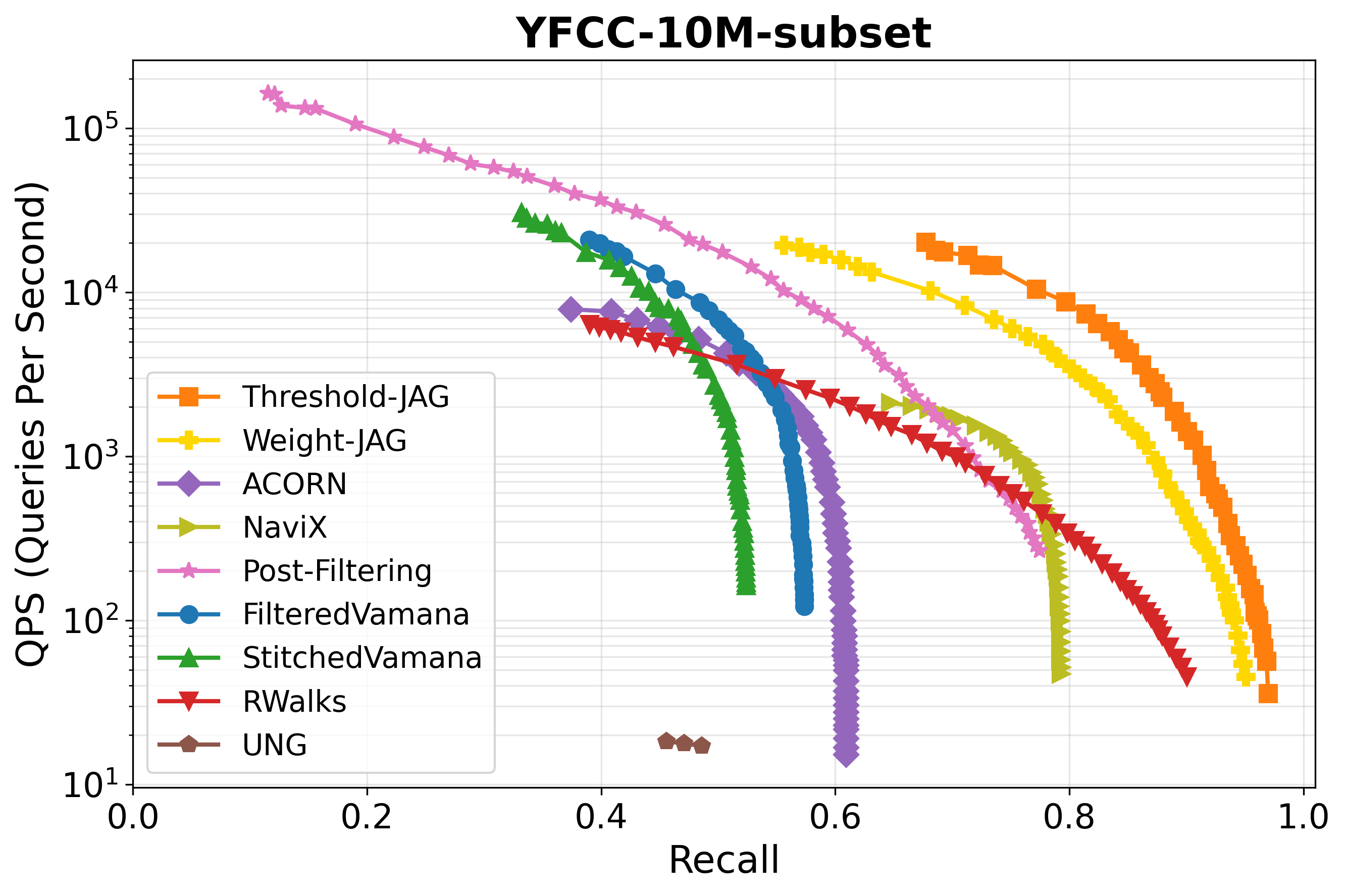}
        \label{fig:yfcc-subset}
    \end{subfigure}
    \vspace{-1em}
    \caption{QPS vs. recall plots for subset filters on the MSTuring-10M, LAION-5M, and 25M datasets, and for boolean filters on the MSTuring-10M dataset. Note that FilteredVamana and StitchedVamana are only for label and subset filters.}
    \label{fig:exp-subset}
    \vspace{-2em}
\end{figure*}

\begin{figure}[!h]
    \centering
    \begin{subfigure}{0.45\textwidth}
        \centering
        \includegraphics[width=\linewidth]{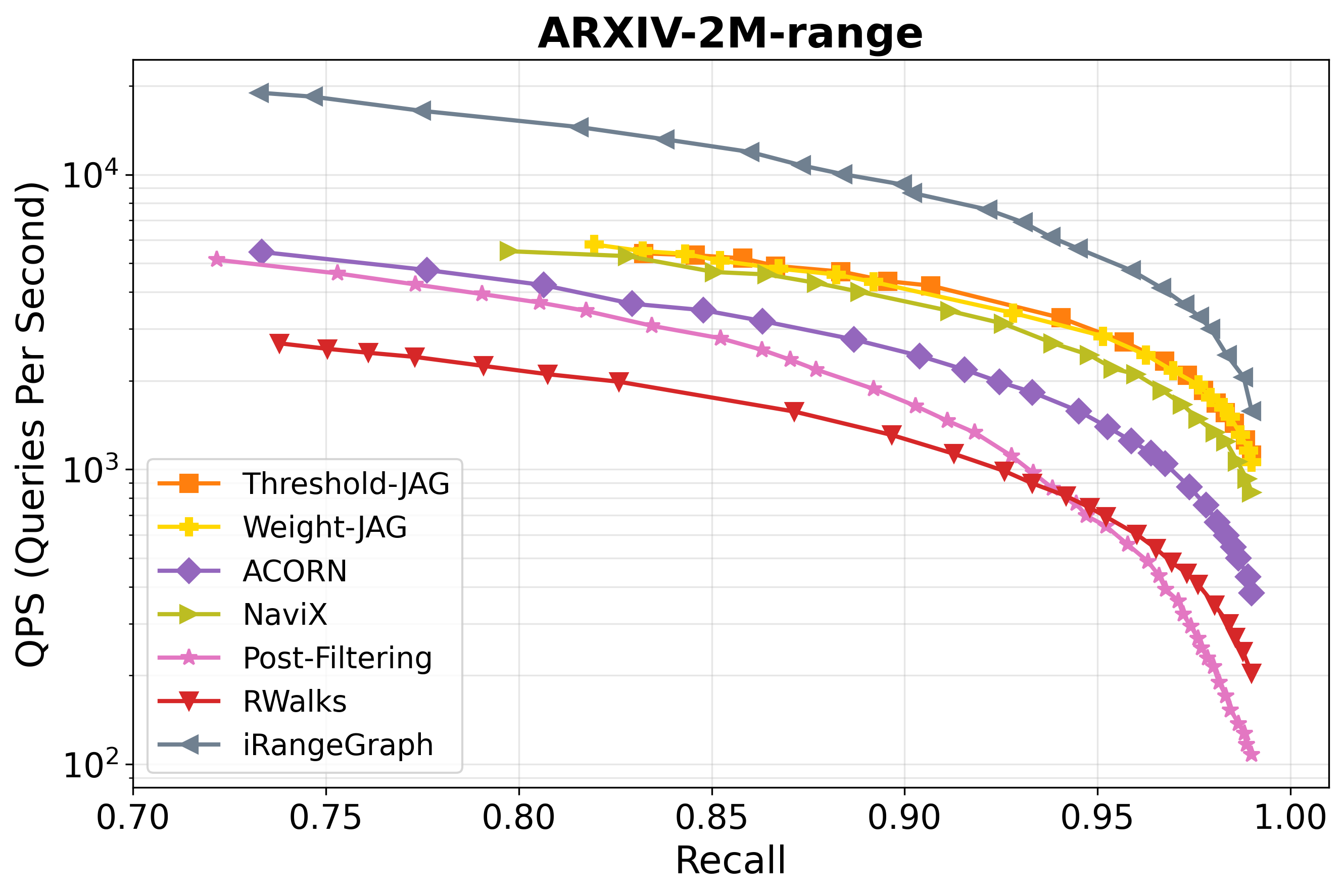}
        \label{fig:arxiv-range}
    \end{subfigure}
    \vspace{-1.5em}
    \\ 
    \begin{subfigure}{0.45\textwidth}
        \centering
        \includegraphics[width=\linewidth]{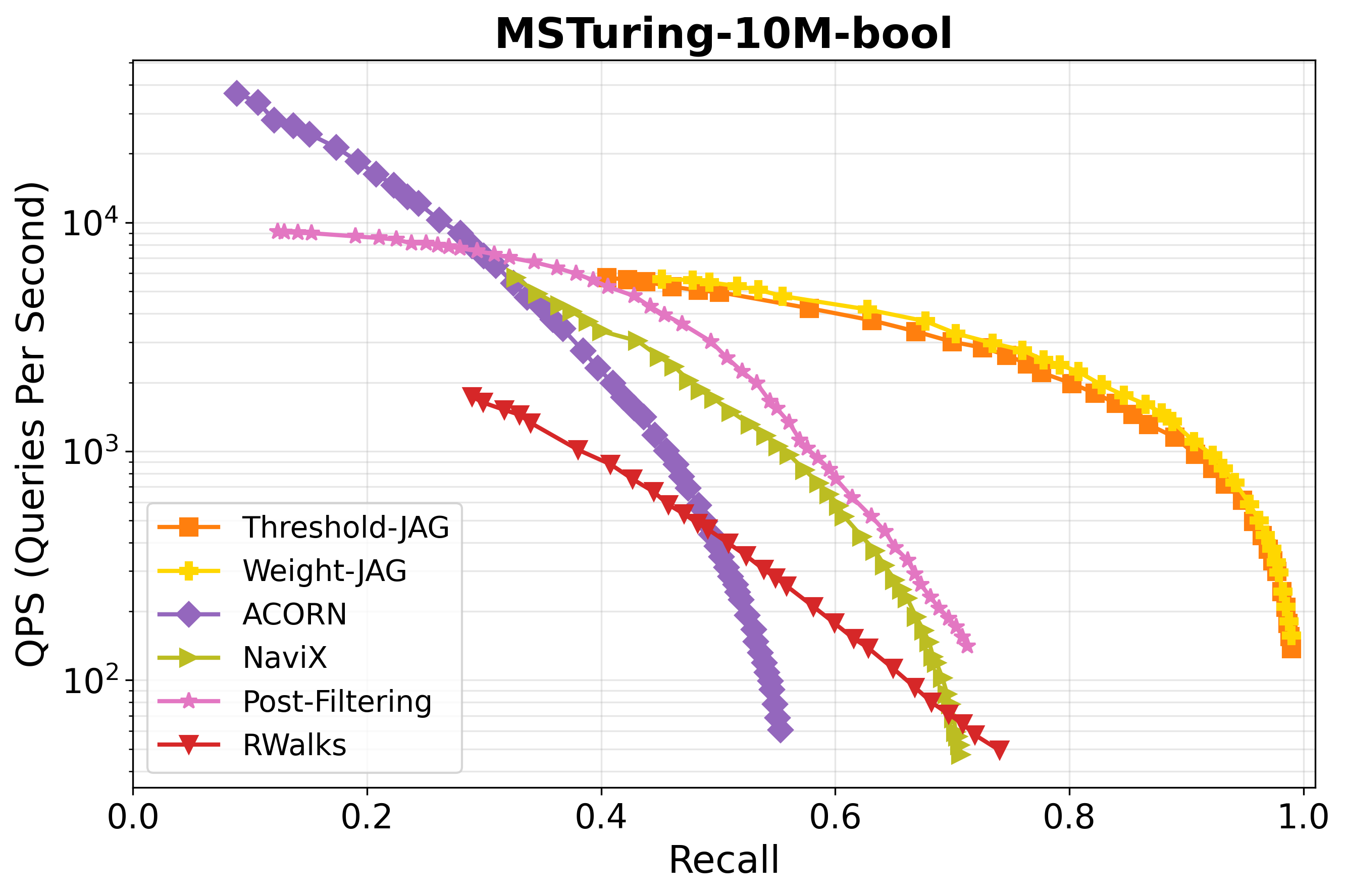}
        \label{fig:msturing-range}
    \end{subfigure}
    \vspace{-1.5em}
    \caption{QPS vs. Recall plot for range filters on the ARXIV and bool filters on MSTuring datasets. Note that iRangeGraph is designed specifically for Range filter}
    \label{fig:exp-range}
\end{figure}

\subsection{Results}

Please see our QPS v.s. recall plots in Figure~\ref{fig:exp-label},~\ref{fig:exp-range},~\ref{fig:exp-subset}, distance computation v.s. recall plots in Figure~\ref{fig:exp-label-dc},~\ref{fig:exp-range-dc},~\ref{fig:exp-bool-dc} in Appendix~\ref{sec:exp-dc}, and indexing time in Table~\ref{tab:indexing-time}. For the Pre-Filtering algorithm, it always achieves perfect recall but its QPS is usually too low to show on the plots. We report its QPS and the number of distance computations it performs in Table~\ref{tab:exp-preprocessing}. In the following, we elaborate on the algorithm analysis for different filters.

\myparagraph{Label filters} Figure~\ref{fig:exp-label} shows the performance of all algorithms on the SIFT1M and ARXIV datasets with label filters.
For the SIFT dataset, all algorithms achieve perfect recall, though with different QPS values.
Our algorithm is slower than those filter specific algorithms like StitchedVamana, FilteredVamana, UNG, but outperforms all other methods.
This is expected, as for non-overlapping label filters, the optimal solution is to build a separate index for each label and then perform standard nearest neighbor search within the subset of points sharing that label — which is essentially what StitchedVamana, FilteredVamana, and UNG are doing for such cases. 
On the ARXIV dataset, our algorithm, along with StitchedVamana, FilteredVamana, and UNG, performs almost identically and better than all other algorithms.

\myparagraph{Range filters} Figure~\ref{fig:exp-msturing-range}, and upper Figure~\ref{fig:exp-range} show the performance of all algorithms on the ARXIV and MSTuring datasets with range filters.
For the ARXIV dataset, all algorithms achieve perfect recall, with our algorithm attaining the highest QPS than all the other filter-agnostic algorithms, but less efficient than iRangeGraph, which is an algorithm specially designed for range queries.
On the MSTuring dataset, it is noteworthy that only our algorithm and iRangeGraph are able to reach perfect recall, while all other general purpose algorithms achieve at most 0.8 recall. 
The reason is that our synthetic range filters contain filter ranges with varying selectivity; we hypothesize that the other algorithms cannot effectively handle highly selective cases.
Please refer to our ablation studies in Figure~\ref{fig:exp-range-sel} for a more detailed analysis.

\myparagraph{Subset filters} Figure~\ref{fig:exp-subset} shows the performance of all algorithms on the MSTuring, YFCC, and LAION datasets with subset filters.
For the MSTuring and YFCC datasets, only our algorithm is able to achieve perfect recall, while all other algorithms reach at most 0.8 on MSTuring and 0.9 on YFCC before their QPS drops below 50.
As noted in \cite{ait2025rwalks}, most queries from the YFCC dataset are extremely selective ($<$1\%), where no previous graph-based algorithm has achieved recall greater than 0.9 without using auxiliary IVF structures or pre-filtering.
Here, we are the first graph-based algorithm to reach almost perfect recall ($>$0.95) with QPS $>$ 50.
Our algorithm performs comparably to StitchedVamana on LAION-5M, and our advantage becomes larger as the dataset size goes to 25M.

\myparagraph{Boolean filters} Lower figure~\ref{fig:exp-range} shows the performance of all algorithms on the MSTuring dataset with boolean filters. Only our algorithm is able to achieve perfect recall, while all other algorithms reach at most 0.8 on MSTuring before their QPS drops below 50.

\input{conclusion}

\input{impact}

\bibliography{example_paper}
\bibliographystyle{icml2026}

\newpage

\appendix

\onecolumn

\input{rel}

\input{full-exp-results}

\input{abl}

\input{full-exp-details}

\end{document}

%% file: conclusion.tex
\section{Conclusion}

In this paper, we introduced \algoname{} (Joint Attribute Graphs), a unified graph-based indexing strategy designed for filtered nearest neighbor search. 
Key to JAG's robustness across different filter types and selectivities are the new notions of attribute distance, filter distance, and capped attribute distances, which enable the use of thresholded distance functions to provide navigational guidance and prevent navigational dead-ends. 
Unlike prior methods that are restricted to specific filter types or struggle to achieve good performance across a broad range of the selectivity regime, \algoname{} unifies edges tailored to different distance thresholds during construction and as a result is capable of handling filters across a broad selectivity range.

%% file: impact.tex
\section*{Impact Statement}

This paper presents work whose goal is to advance the field of machine learning. There are many potential societal consequences of our work, none of which we feel must be specifically highlighted here.

%% file: rel.tex
\section{Related Work}

One solution to filtered vector search is to simply leverage the database to handle the filter.
\emph{Pre-filtering} first isolates the subset of points satisfying the metadata constraints (e.g., using a SQL query) and subsequently computes distances within this reduced candidate set. 
While this method guarantees correctness, it becomes computationally expensive in high-selectivity regimes where the filtered subset remains large.

\myparagraph{Fully-Oblivious Methods}
We define fully-oblivious methods as those that ignore attributes at index construction time.
\emph{Post-filtering} generates an initial list of nearest neighbors using a standard unfiltered index (e.g., HNSW~\cite{HNSW16} or Vamana~\cite{DiskANN19}) and retrospectively discards items that fail the filter criteria. This approach is effective when filters are permissive (high selectivity) but often fails to retrieve sufficient valid neighbors when the valid items are sparse or distant from the query in the vector space.

To mitigate the recall issues of standard post-filtering, more advanced fully-oblivious approaches such as ACORN~\cite{acorn} and NaviX~\cite{sehgal2025navix} utilize indexes constructed solely from vector data but modify the search procedure. 
These methods typically employ expanded neighborhood strategies—such as two-hop traversal—to explore a larger region of the vector space, aiming to encounter a sufficient number of valid candidates. 
However, these approaches implicitly assume that points satisfying the filter are uniformly distributed within the vector neighborhood. 
Consequently, performance degrades significantly when the filter is negatively correlated with vector distance or when selectivity is low. 

\myparagraph{Filter-Aware Methods} Filter-aware algorithms tailor the index structure to specific classes of query predicates known a priori. For example, FilteredVamana and StitchedVamana~\cite{gollapudi2023filtered} modify the DiskANN architecture to accommodate label constraints: the former restricts navigation to valid points, while the latter merges separate graphs constructed for individual filters. 
Similarly, other specialized methods augment proximity graphs with auxiliary data structures, such as tries or segment trees, to handle specific filter types. 
While these solutions excel at their targeted tasks (e.g., label equality or range search), they lack generalization; optimizing for one filter type often precludes efficient handling of others, such as Boolean or Subset queries.

\myparagraph{Attribute-Aware (Filter-Agnostic) Methods} 
Attribute aware (or filter-agnostic) methods leverage the attributes at index construction time to build indexes that are optimized for many (unknown) queries.
There are relatively few methods that can be categorized as attribute aware.
NHQ~\cite{NHQ} is an early filtered search algorithm that  defines an attribute distance between two points based solely on whether their attributes are equal, and combines this with the standard vector distance using a weighted average.  As a result, NHQ can only handle equality-based filters (and thus it could fairly be classified as a filter-aware method).
The only other attribute-aware method, to the best of our knowledge, is a recent work called RWalks~\cite{ait2025rwalks} which uses a standard
(unfiltered) index and augments it for filtered search. 
At index construction time, the algorithm performs a random walk on the unfiltered index to compute an aggregated attribute vector.
It then defines the filter distance as the fraction of labels in this vector that match the query filter.
At query time, this filter distance is combined with the usual vector distance using a weighted scoring function to guide greedy search.

%% file: full-exp-results.tex
\section{Additional Experimental Results}

\begin{figure}[h]
    \centering
    \includegraphics[width=0.43\textwidth]{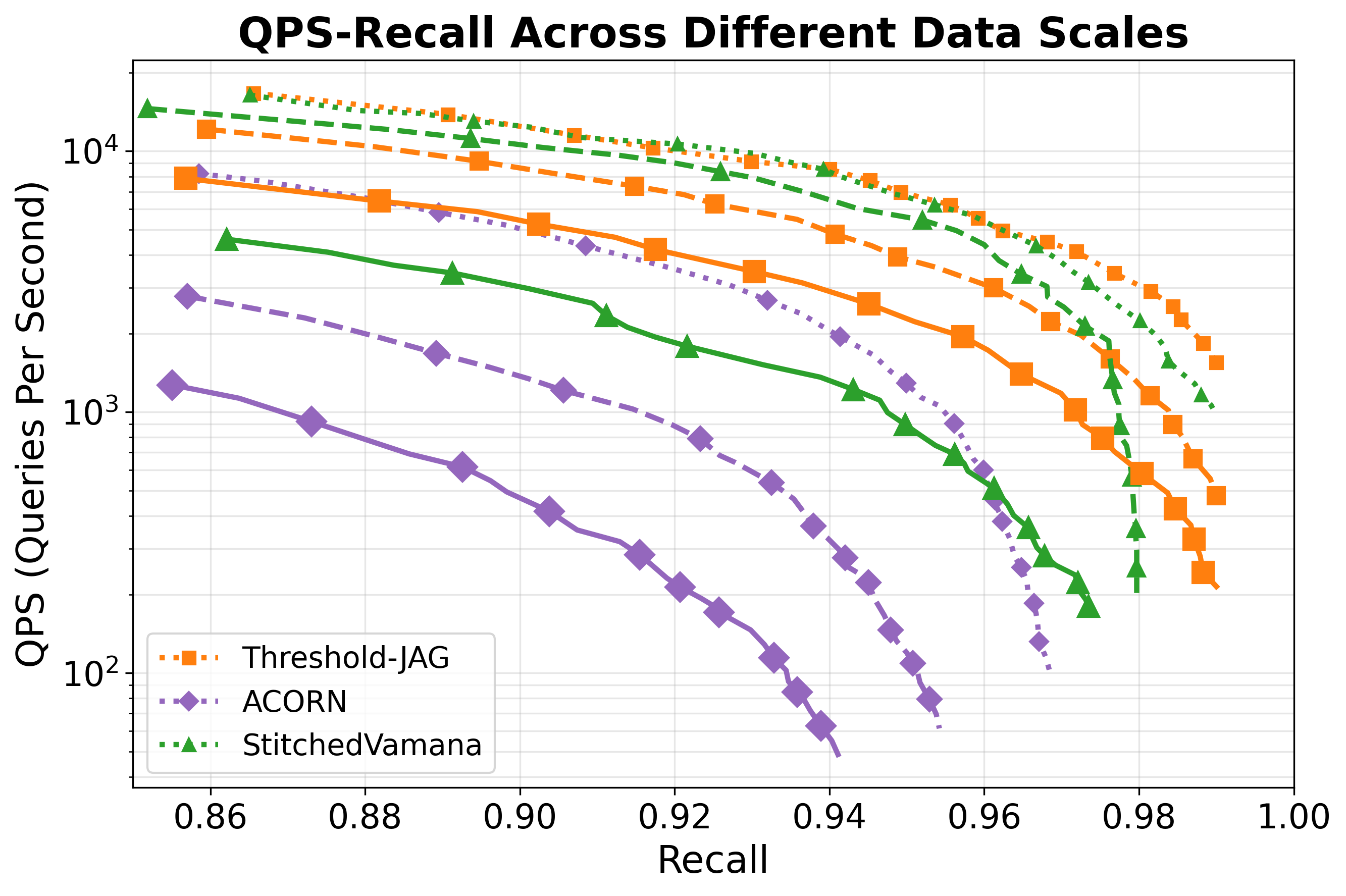}
    \caption{\small QPS vs. recall for subset filters on the LAION dataset. Each algorithm is represented by a different color; different line styles denote different dataset sizes: dotted for 1M, dashed for 5M, and solid for 25M.}
    \label{fig:exp-scaling}
\end{figure}

\myparagraph{Scaling Experiment}
For the LAION dataset, we select the three best algorithms—ours, ACORN, and StitchedVamana—and run them on datasets of different sizes (1M, 5M, and 25M) to evaluate their scaling behavior.
Please see Figure~\ref{fig:exp-scaling} for the results. Our algorithm performs comparable with StitchedVamana on 1M and 5M scales, and then our advantage becomes more obvious in the 25M scale.
All the QPS curves move downward as the dataset size increases from 1M to 5M to 25M, while our algorithm continues to achieve perfect recall and maintains QPS $>$ 50.

\begin{figure}
    \centering
    \begin{subfigure}[t]{0.45\textwidth}
        \centering
        \includegraphics[width=\linewidth]{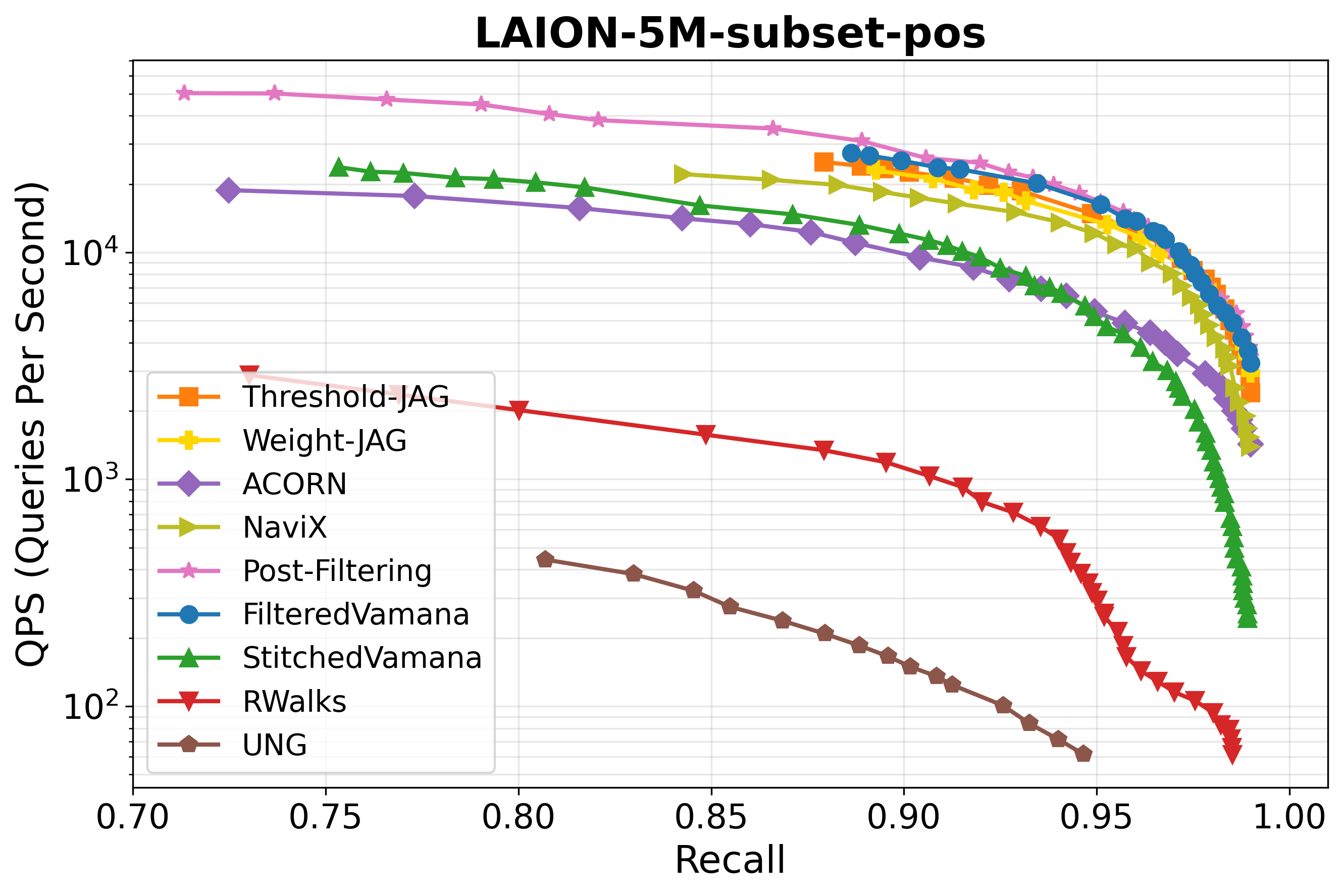}
        \label{fig:laion-5m-subset-pos}
    \end{subfigure}
    \begin{subfigure}[t]{0.45\textwidth}
        \centering
        \includegraphics[width=\linewidth]{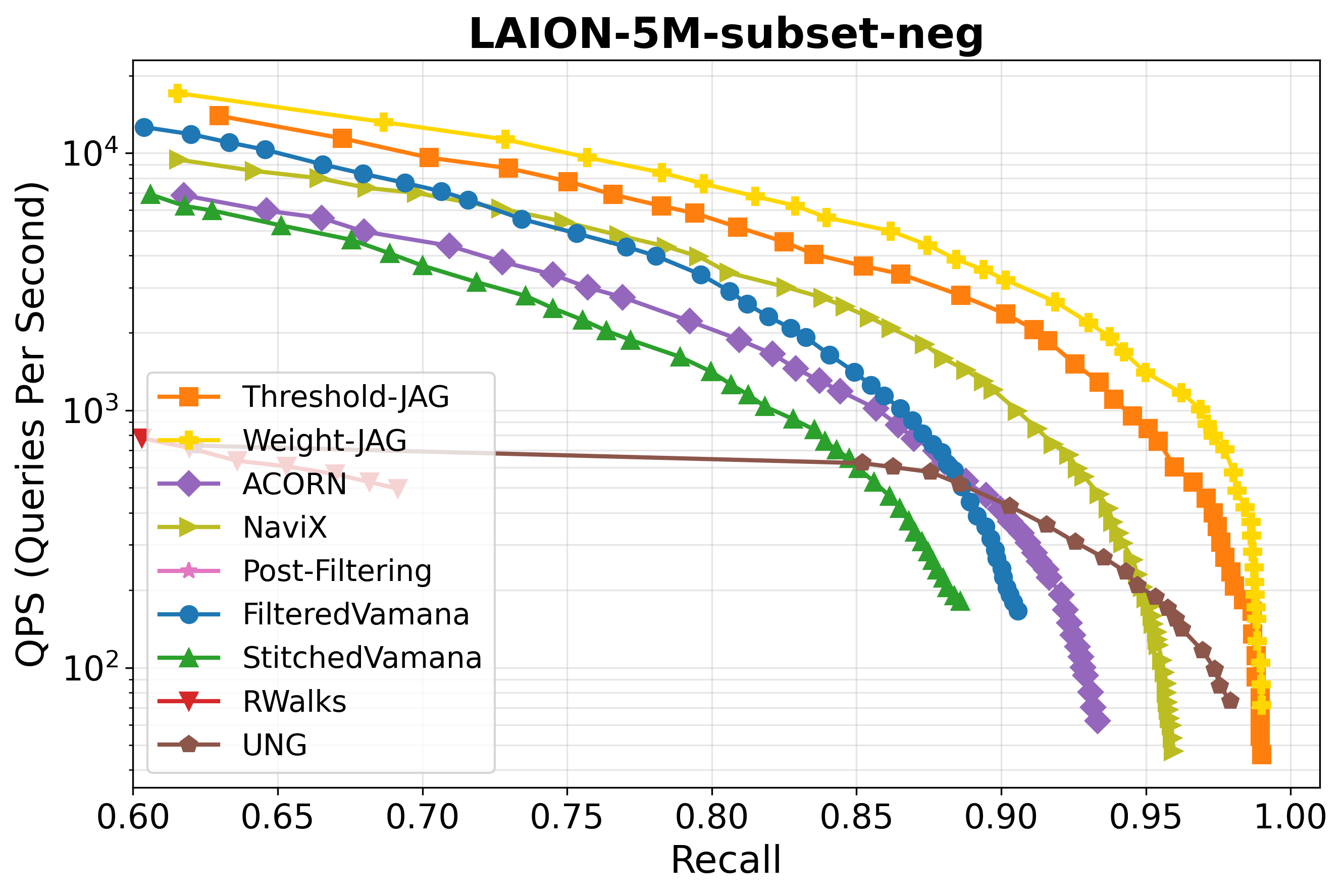}
        \label{fig:laion-5m-subset-neg}
    \end{subfigure}
    \vspace{-1em}
    \caption{QPS vs. recall plots for subset filters with positive and negative correlations on the LAION-5M dataset.}
    \label{fig:exp-correlation}
\end{figure}

\myparagraph{Correlation}
We test how the algorithms perform when the query filter is spatially correlated with the vectors.
For the LAION-5M dataset, each keyword represents a point cluster in the vector space.
To create a positive correlation between the query and the filter, we set the query filter to be the closest keyword cluster to the query image embedding.
To create a negative correlation, we pick the query filter to be the farthest keyword cluster from the query image embedding.
Please see Figure~\ref{fig:exp-correlation} for the results.
For the positive correlation case, our algorithm, FilteredVamana, Post-Filtering, and NaviX perform almost equally well.
Interestingly, Post-Filtering performs poorly on other subset filter cases but is extremely fast in this scenario.
This is because the way we select query filters causes significant overlap between the query’s neighborhood and the filtered subset—meaning the points closest to the query vector are highly likely to satisfy the filter.
For the negative correlation case, our algorithm, NaviX, and UNG are able to achieve $>$ 0.95 recall, while our algorithm has the highest QPS.

\myparagraph{Varying Filter Selectivity}
We evaluate how the algorithms perform under different query selectivities. 
We conduct the experiment on MSTuring-10M using range filters and 
separately measure the recall of each algorithm when the query filter selectivities are at $1, 10^{-1}, 10^{-2}, 10^{-3}, 10^{-4}, 10^{-5}$. 
For each selectivity level, we report the maximum recall achieved while maintaining
QPS $> 5000$.

\begin{figure}[h]
    \centering
    \includegraphics[width=0.45\textwidth]{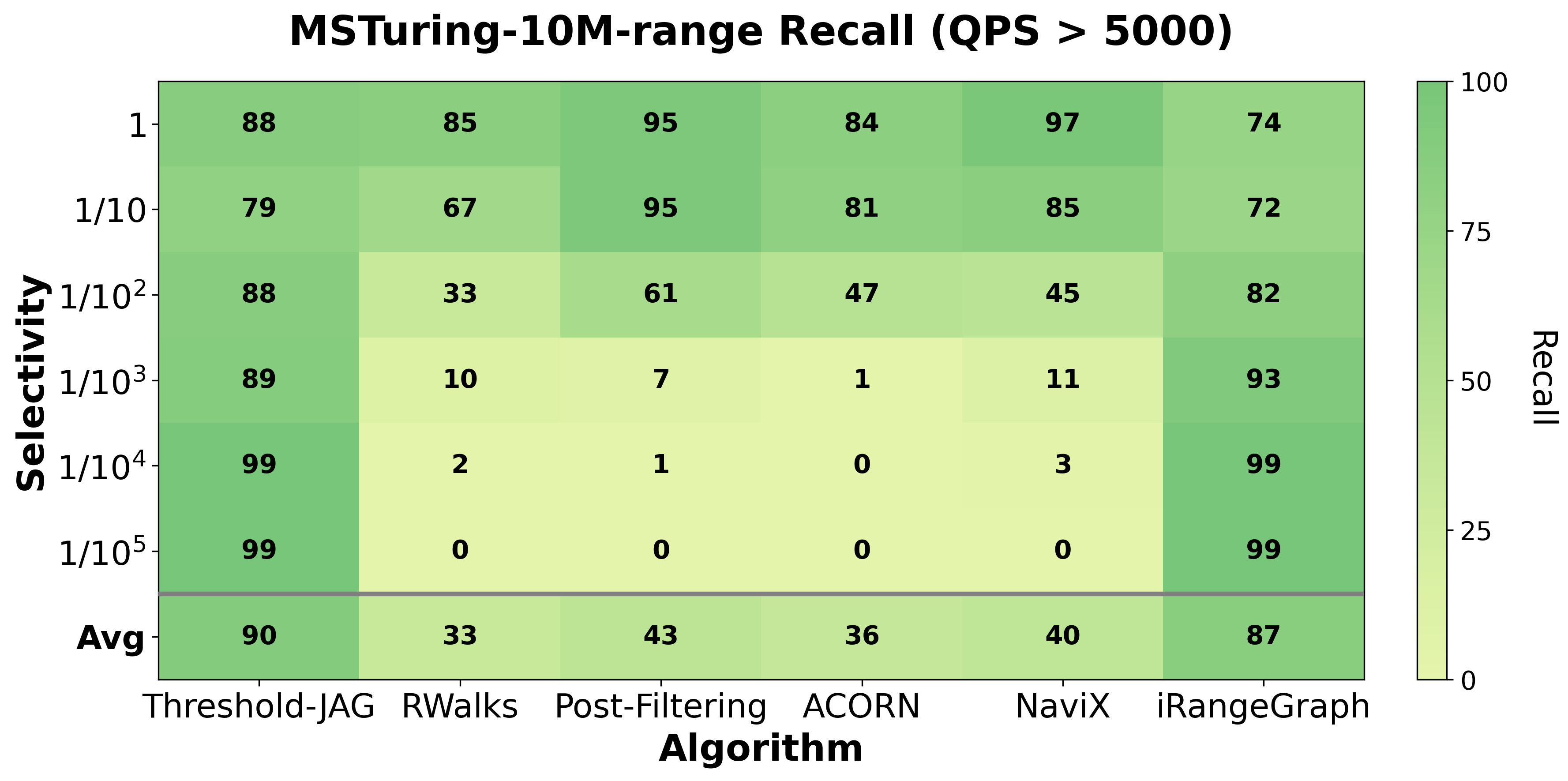}
    \caption{Recall for range filters on the MSTuring-10M dataset with different filter selectivities when maintaining QPS $>$ 5000.}
    \label{fig:exp-range-sel}
\end{figure}

As shown in Figure~\ref{fig:exp-range-sel}, all general filtered vector search
algorithms perform well only at high selectivity, and their recall gradually drops 
to zero as selectivity decreases. In contrast, our \algoname{} algorithm is 
comparable to iRangeGraph, which is specifically designed for range filters.

%% file: abl.tex
\section{Ablation Studies}
\label{sec:ablation}

\begin{figure*}[t]
    \begin{subfigure}[t]{0.45\textwidth}
        \centering
        \includegraphics[width=\linewidth]{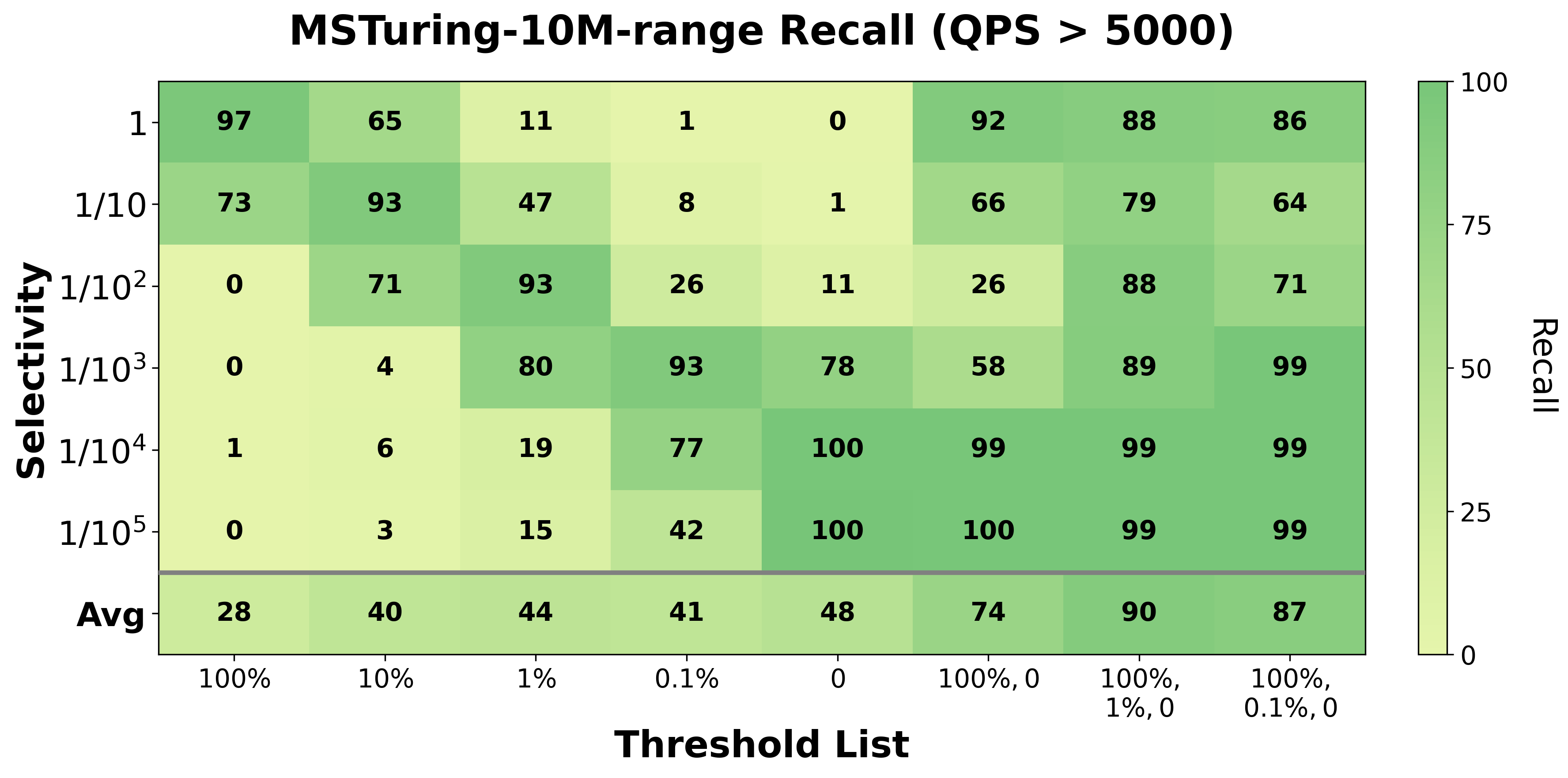}
        \label{fig:ablation-lexical-range-msturing}
    \end{subfigure}
    \begin{subfigure}[t]{0.45\textwidth}
        \centering
        \includegraphics[width=\linewidth]{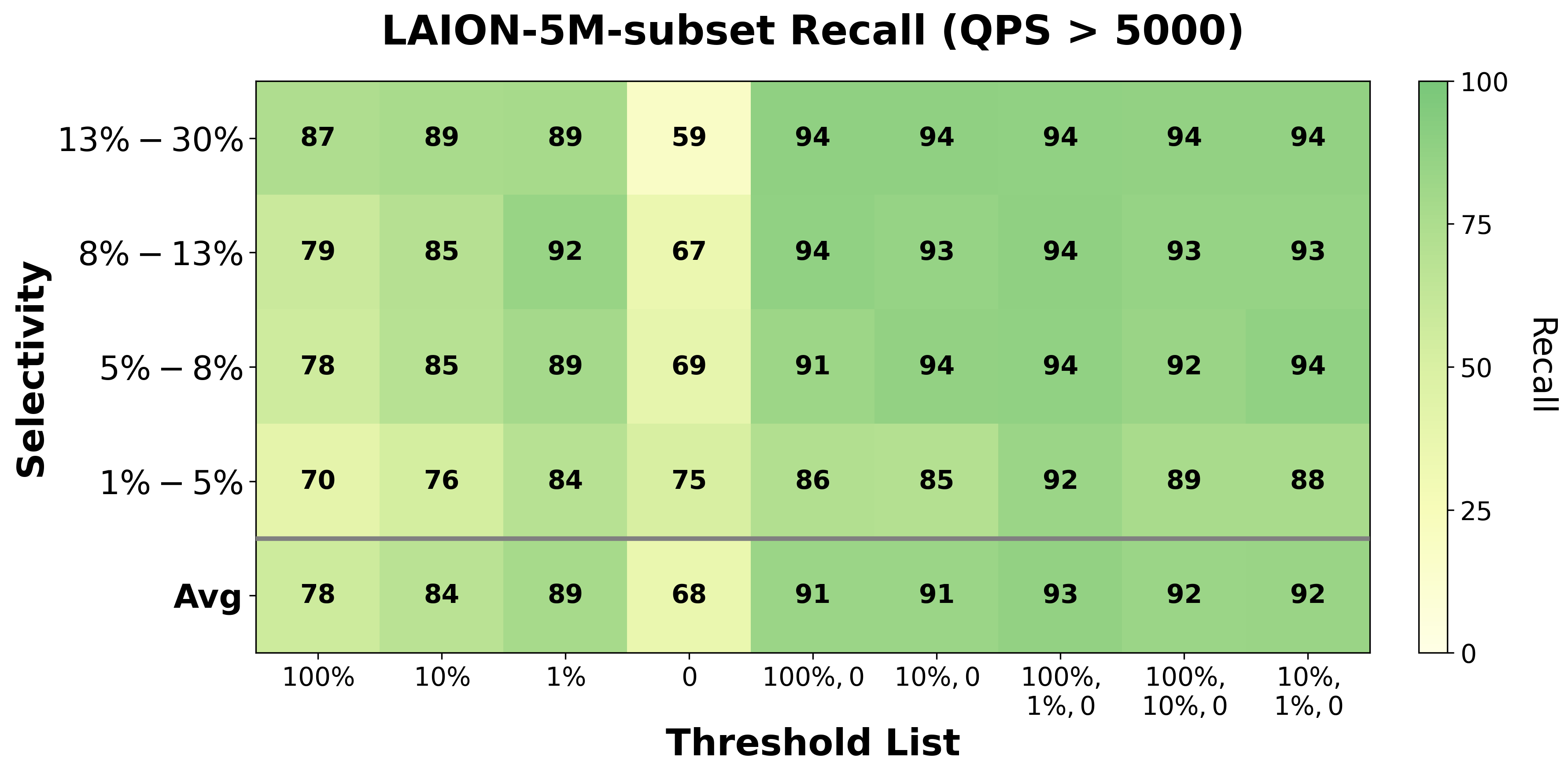}
        \label{fig:ablation-lexical-subset-laion5m}
    \end{subfigure}
    \centering
    \begin{subfigure}[t]{0.45\textwidth}
        \centering
        \includegraphics[width=\linewidth]{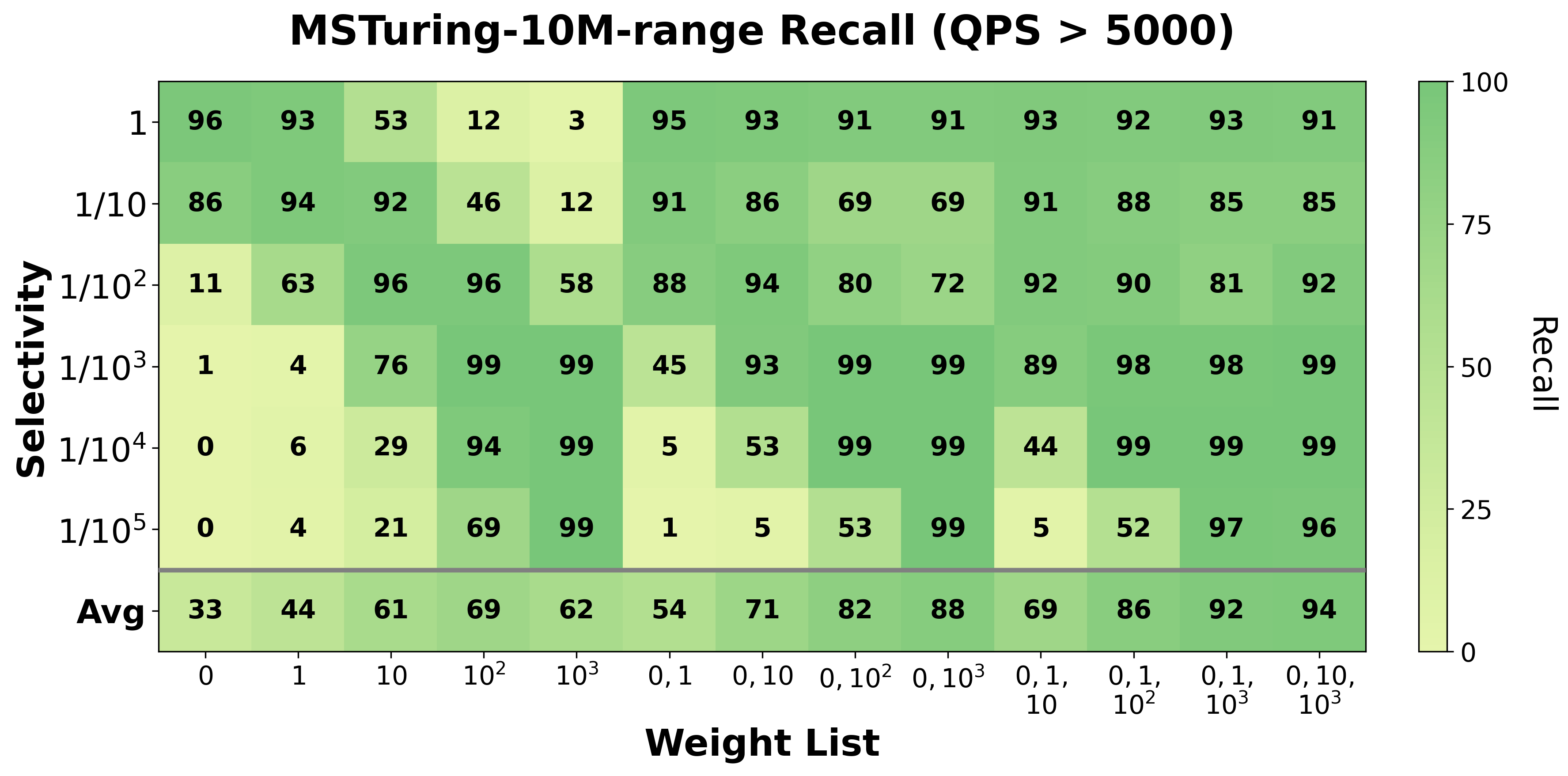}
        \label{fig:ablation-weight-range-msturing}
    \end{subfigure}
    \begin{subfigure}[t]{0.45\textwidth}
        \centering
        \includegraphics[width=\linewidth]{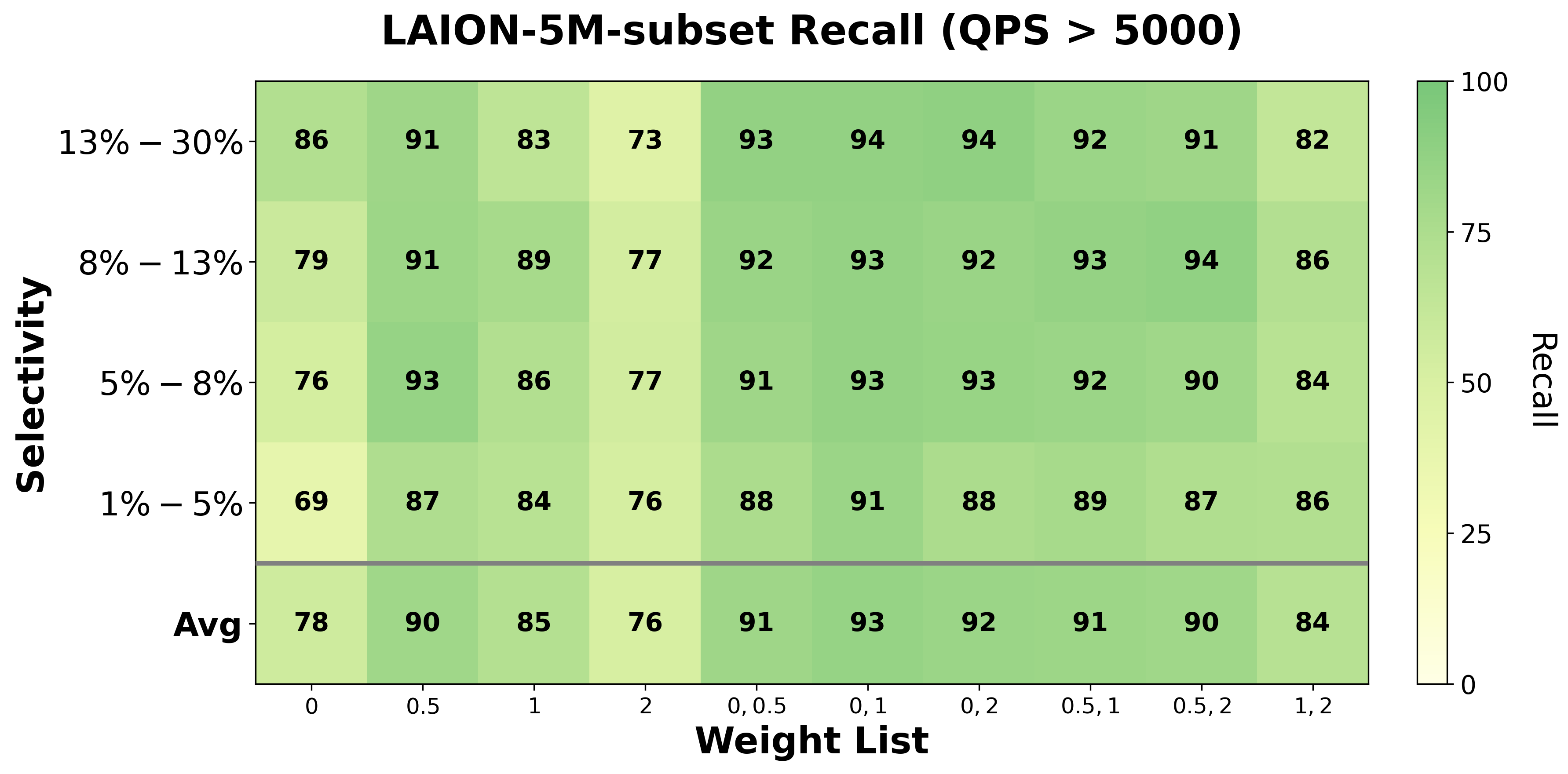}
        \label{fig:ablation-weight-subset-laion5m}
    \end{subfigure}
    \vspace{-1em}
    \caption{Query recall for different filter selectivities using different thresholds on the MSTuring-10M dataset with range filters (upper left) and the LAION-5M dataset with subset filters (upper right), and using different weights on the MSTuring-10M dataset with range filters (lower left) and the LAION-5M dataset with subset filters (lower right)}
    \label{fig:ablation-table}
\end{figure*}

\myparagraph{Multiple thresholds}
We perform an ablation study on MSTuring-10M for range filters and on LAION-5M for subset filters to examine how different thresholds help our algorithm handle filters with varying selectivity. Note that the original query filters in Section~\ref{sec:experiments} consist of mixed selectivities. Here, we evaluate the performance of building our algorithm with different single thresholds and measure how they behave under filters of varying selectivity levels. 

For MSTuring-10M-range, we consider six query selectivities: $1, 10^{-1}, 10^{-2}, 10^{-3}, 10^{-4}, 10^{-5}$. For LAION-5M-subset, we sort all query selectivities and evenly divide them into four ranges ($13\%-30\%$, $8\%-13\%$, $5\%-8\%$, $1\%-5\%$). We use the same degree limit for all threshold choices. Please refer to the two upper plots in Figure~\ref{fig:ablation-table} for the results.

When the single filter threshold is set to 0\%—meaning we strictly compare by filter distance and only use standard distance when the filter distance is equal—the resulting index achieves perfect recall in low-selectivity cases (for both filter types). However, its recall gradually decreases as the selectivity increases. This is because, as selectivity weakens, the filter becomes less important, yet the query comparison still overemphasizes filter distance.

The opposite occurs with a 100\% threshold, which is equivalent to searching purely by standard distance. This index performs well in high-selectivity cases but worse as the selectivity becomes more challenging. Interestingly, intermediate thresholds each perform best within a specific range of the selectivity spectrum and degrade as the selectivity deviates from their “comfort zone.” For example, in the subset-filter case, the 1\% threshold performs best on 
queries whose selectivities are between $8\%-13\%$ (i.e., the 
top 25--50\% after ordering queries by selectivity). Its performance drops 
when the selectivity becomes either higher or lower. Similarly, in the range case, the 1\% threshold performs best at $1/100$ and decreases in both directions.

By merging three different thresholds $\{100\%,1\%,0\}$ (sharing the same degree bound), the final index achieves strong and stable performance across all query selectivity levels.

\paragraph{Weight-\algoname}
We also implement a weighted variant of \algoname, in which attribute 
distance and vector distance are combined using different weights. 
For a weight $w$, we define
\[
D_A^w(u,v) = w \cdot \mathrm{dist}_A(a_u, a_v) + \mathrm{dist}(x_u, x_v).
\]
We build the index graph using a set of such weights. Similar to Threshold-\algoname, when calling \textsf{Insert} and \textsf{JointRobustPrune}, we iterate over the weight list and use $D^w_A(u,v)$ as the comparator.

During query processing, 
we still compare $\mathrm{dist}_F(\cdot)$ first, followed by the standard vector 
distance. We refer to this variant as Weight-\algoname.

\myparagraph{Multiple weights}
We also conduct an ablation study on Weight-\algoname{} to examine how 
different weights help handle filters with varying selectivities. From the 
lower two plots in Figure~\ref{fig:ablation-table}, we observe that Weight~0, 
which corresponds to building the proximity graph purely using the standard 
vector distance, achieves the best recall for standard nearest neighbor 
queries (selectivity 1). As the weight increases, the resulting proximity 
graph becomes more tailored to low-selectivity queries. By merging graphs 
constructed with different weights, the final index achieves strong recall 
across all query selectivity levels.

\myparagraph{Comparing Threshold-\algoname{} and Weight-\algoname{}}
Though both implementations of our \algoname{} (Threshold and Weight) can handle general filter search with different selectivities, they have different advantages. The Weight-\algoname{} may have slightly better recall performance on MSTuring-10M-range, but its weight choices are not quite robust. For example, the optimal weight combination on MSTuring-10M-range is $\{0,10,1000\}$, but the optimal combination for LAION-5M-subset is $\{0,2\}$. On the contrary, for Threshold-\algoname, we choose the thresholds from the a subset of $\{100\%,10\%,1\%,0.1\%,0\}$ throughout our experiment.

To summarize our experimental findings on these two variations on our algorithm, we find that Threshold-\algoname{} performs consistently well across different datasets and requires less tuning. 
On the other hand, Weight-\algoname{} occasionally obtains the best results by a small margin; however, it is less robust across different datasets and requires more tuning. 
Our recommendation is to use Threshold-\algoname{} as a robust and good-quality default.

\myparagraph{Unifying multiple graphs} A natural question to ask is: \emph{if the query selectivity were known in advance, could we simply build multiple independent indices, each tailored to a specific selectivity range, and route each query to the corresponding index?} Our results suggest that this strategy would be suboptimal.

From the LAION-5M experiment in Figure~\ref{fig:ablation-table}, we observe that for the queries with selectivities from $1\%-5\%$, the threshold combination $\{100\%, 1\%, 0\}$ outperforms \emph{every} individual threshold. Similarly, for Weight-\algoname, the weight combination $\{0, 2\}$ outperforms any individual weight configuration.

These findings indicate that greedy search benefits from access to edges constructed under \emph{multiple} thresholds or weights. The search is able to adaptively traverse these heterogeneous edges to reach the target efficiently---something that a single, selectivity-specific index cannot provide.

%% file: full-exp-details.tex
\section{Experimental Setup and Details}\label{sec:full-exp-details}

\subsection{Machines}
All experiments are conducted on a Google Cloud VM with the following specifications: an n2-highmem-64 instance equipped with 32 visible Intel Ice Lake vCPUs (1 vCPU per physical core) and 512 GB RAM. We run our experiments using 32 threads.

\subsection{Additional Dataset Details}\label{sec:full-dataset}

\begin{table*}[]
\small
\setlength{\tabcolsep}{4pt}
\caption{Dataset sizes and average selectivities. QPS (queries per second) and DC (distance computations) for Pre-Filtering algorithm.}
\label{tab:exp-preprocessing}
\resizebox{\textwidth}{!}{
\begin{tabular}{cccccccccccc}
\toprule
&
&
\makecell{SIFT1M\\Label} &
\makecell{ARXIV\\Label} &
\makecell{ARXIV\\Range} &
\makecell{LAION-1M\\Subset} &
\makecell{LAION-5M\\Subset} &
\makecell{LAION-25M\\Subset} &
\makecell{YFCC\\Subset} &
\makecell{MSTuring10M\\Subset} &
\makecell{MSTuring10M\\Range} &
\makecell{MSTuring10M\\Bool} \\
\midrule
\multirow{2}{*}{Dataset} & Size & $10^6$ & $2.7\cdot 10^6$ & $2.7\cdot 10^6$ & $10^6$ & $5\cdot 10^6$ & $2.5\cdot 10^7$ & $10^7$ & $10^7$ & $10^7$ & $10^7$ \\
 & \makecell{Avg\\Selectivity} & 0.083 & 0.37 & 0.43 & 0.099 & 0.098 & 0.099 & 0.018 & 0.15 & 0.17 & 0.15 \\
\midrule
\multirow{2}{*}{\makecell{Pre\\Filtering}} & QPS             & 412   & 5.4  & 4.6  & 146   & 28    & 5.6   & 5.2  & 27   & 31   & 2.3  \\
& DC & $8.3\cdot10^4$ & $1.0\cdot10^6$ & $1.1\cdot10^6$ &
                   $9.9\cdot10^4$ & $4.8\cdot10^5$ & $2.4\cdot10^6$ &
                   $1.8\cdot10^5$ & $1.5\cdot10^6$ & $1.7\cdot10^6$ & $1.5\cdot10^6$ \\
\bottomrule
\end{tabular}}
\end{table*}

\smallskip
\noindent\textbf{SIFT.}
The SIFT dataset~\cite{lowe2004distinctive,jegou2008hamming} is one of the standard benchmarks for approximate nearest neighbor (ANN) and filtered-ANN research. 
It consists of one million points in $\mathbb{R}^{128}$. 
Following~\cite{acorn,cai2024navigating-ung}, we assign each point an integer attribute uniformly sampled from $\{1, \ldots, 12\}$, and each query is assigned a label filter.
A point satisfies the filter if its attribute matches the query’s label.

\smallskip
\noindent\textbf{ARXIV.}
We use the dataset and filter setup from the recent benchmark in~\cite{arxiv-benchmark}, which contains text embeddings of 2.7M arXiv papers generated using the \texttt{stella\_en\_400M\_v5} model~\cite{zhang2024jasper-stella}. 
We apply both range and label filters:  
for range filters, the attribute is the publication date, and the query specifies a target time range;  
for label filters, the attribute is the number of subcategories assigned to each paper, and the query retrieves neighbors with a given number of subcategories.

\smallskip
\noindent\textbf{LAION.}
We use the LAION dataset adopted in~\cite{acorn,cai2024navigating-ung}, which consists of CLIP embeddings of web-scraped images.
Following the setup in~\cite{acorn}, we construct a vocabulary of 30 common keywords and assign to each image the three keywords whose embeddings are closest to that image vector.
Each query specifies one keyword as the filter, and the search retrieves image vectors whose assigned keyword sets contain the specified term.
We instantiate this setup at three scales—1M, 5M, and 25M points.
Because the keywords capture the semantic content of images, we vary the correlation between filter and vector similarity by selecting different types of query filters:
\emph{positive} (the keyword closest to the query image),
\emph{random} (a randomly chosen keyword), and
\emph{negative} (the keyword farthest from the query image embedding).

\smallskip
\noindent\textbf{YFCC.}
The YFCC dataset is the standard filtered-ANN benchmark used in the filtered search track of the NeurIPS 2023 competition.
It consists of CLIP embeddings of 10M randomly selected images from YFCC100M (Flickr)~\cite{thomee2016yfcc100m}.
Each image is associated with a ``bag of tags'' extracted from metadata such as the description text, camera model, year, and country, totaling over 200K unique attributes.
Each query specifies a set of tags as the filter, and the goal is to retrieve images whose tag sets contain all of the query’s filter terms.

\smallskip
\noindent\textbf{MSTuring.}
The MSTuring dataset consists of Bing search queries and corresponding answers. 
We use a 10M-point subset with 100-dimensional embeddings and synthetically construct \emph{subset}, \emph{range}, and \emph{boolean} filters with mixed selectivity.  

\emph{Subset filters:} we define 30 binary attributes; each point independently includes each attribute with probability $1/2$. 
For a query, we randomly select $k \in \{0,2,4,6,8,10,12,14,16\}$ attributes and require that candidate points contain all selected attributes.  

\emph{Range filters:} each point is assigned an integer attribute in $[0,10^6]$, and queries specify random intervals of length $10^6/k$ for $k \in \{1,10,100,1000,10^4,10^5\}$ to control selectivity.  

\emph{Boolean filters:} Each query filter is a random Boolean function $f$ over 15 Boolean variables, and each point’s attribute is a random instantiation $x$ of these variables.
A point satisfies the filter if $f(x) = \texttt{true}$.
We control the selectivity by generating functions whose pass rates fall into four ranges: $(1/2^4,1)$, $(1/2^8, 1/2^4)$, $(1/2^{12}, 1/2^8)$, and $(0,1/2^{12})$.

\subsection{Implementation Details for \algoname}\label{sec:implementation_details}

For Threshold-\algoname{}, in practice, we determine the thresholds of Threshold-\algoname{} by selecting the threshold values at specific points. 
For each point $p$ with attribute $a_p$, we sample a set $V$ of points (e.g., $|V| = 500$), compute the distribution of $dist_A(a_p, a_V)$, and consider candidate thresholds from the set of quantiles $\{100\%, \allowbreak 10\%, 1\%, 0\%\}$. 
We then choose the threshold list that yields the best performance.

For Weight-\algoname{}, in practice, for each point $p$ with attribute $a_p$, we sample a set $V$ of points, 
compute the standard deviation $\sigma_A$ of the attribute distance 
$\mathrm{dist}_A(a_p, a_V)$ and the standard deviation $\sigma$ of the vector 
distance $\mathrm{dist}(x_p, x_V)$. Let $h = \sigma / \sigma_A$. 
We then select weights from the set 
\[
\{0,\, h,\, 2h,\, 5h,\, 10h,\, 100h,\, 1000h\}
\]
and choose the subset that yields the best performance. Note that, compared to Threshold-\algoname, the weight range used in 
Weight-\algoname{} is much larger, and some extreme values (e.g., $1000h$) 
turn out to be useful in our experiments. This indicates that 
Weight-\algoname{} is less robust than Threshold-\algoname. 
Since using weighted combinations of attribute and vector distances has 
been explored in prior work \cite{NHQ,ait2025rwalks}, we include this variant 
primarily for ablation studies. For both performance and robustness, 
we recommend the Threshold-\algoname{} version.

On Line 8 of Algorithm~\ref{alg:pruning}, we let $V'_t$ include $v$ if $v \in V'$ was already added by a previous threshold $t$. This is done to allow better pruning and improved edge utilization.

To accelerate the index building process without triggering too many JointRobustPrune calls, we introduce an early exit condition on Line 10 of Algorithm~\ref{alg:pruning} where we terminate when $|V'_t| \ge 0.9 \cdot \text{deg} / |T|$. This means we only store $90\%$ of the degree limit after JointRobustPrune, with the expectation that this will prevent JointRobustPrune from being triggered every time a new edge is added. This is a common technique used in previous implementations.

For the YFCC and LAION datasets, we adopt a modification to the attribute distance design due to the high variance in their label distributions. For example, on YFCC, this variance is characterized by two factors:
\begin{enumerate}
    \item The dataset contains $L=200363$ distinct labels, and a single data point can be associated with between $0$ and $1517$ labels.
    \item There is a large disparity in attribute frequency: some labels are shared across more than $30\%$ of the dataset, while others appear only once.
\end{enumerate}
To mitigate this imbalance and prioritize less frequent attributes, we gather the frequency $p_i$ for each attribute $i\in [L]$ and weight them by $\log(1/p_i)$. We then define the attribute distance as:
$$dist_A(a_u,a_v)= C - \sum\limits_{i\in a_u\cap a_v}\log(1/p_i)$$
for a sufficiently large constant $C$. This design prioritizes filters appearing less frequently in the dataset and focuses on the shared labels, rather than the differences between them.

\subsection{Additional Algorithm Details}\label{sec:full-exp-alg}

Below, we briefly describe each baseline algorithm
we evaluate and how we choose its parameters.

Specifically, if an algorithm and dataset were previously evaluated in the
original paper, we directly adopt their parameter settings. Otherwise, we refer
to the parameter options recommended in the original paper and select the
configuration that achieves the highest QPS at recall~0.95. If the optimal recall the algorithm achieves at QPS=50 is less than 0.95 we then pick the configuration that achieves the highest possible recall.

\begin{table*}[t]
\centering
\renewcommand{\arraystretch}{1.2}
\vspace{-1em}
\setlength{\tabcolsep}{4pt}
\caption{\small Compatibility of algorithms with different filter types. 
\cmark\ indicates support and \xmark\ indicates lack of support.}
\label{tab:alg-filter}
\begin{tabular}{
    c|
    >{\centering\arraybackslash}m{1.3cm}
    >{\centering\arraybackslash}m{1.7cm}
    >{\centering\arraybackslash}m{1.3cm}
    >{\centering\arraybackslash}m{1.3cm}
    >{\centering\arraybackslash}m{1.3cm}|
    >{\centering\arraybackslash}m{1.3cm}
    >{\centering\arraybackslash}m{1.3cm}
    >{\centering\arraybackslash}m{1.3cm}
    >{\centering\arraybackslash}m{1.3cm}
    >{\centering\arraybackslash}m{1.3cm}
}
\toprule
\makecell{\textbf{Filter}\\\textbf{Type}} &
\textbf{\algoname} &
\makecell{\textbf{Post}\\\textbf{Filtering}} &
\textbf{ACORN} &
\textbf{NaviX} &
\makecell[c]{\textbf{RWalks}\footnotemark} &
\makecell{\textbf{Filtered}\\\textbf{Vamana}} &
\makecell{\textbf{Stitched}\\\textbf{Vamana}} &
\textbf{NHQ} & \textbf{UNG} &
\makecell{\textbf{iRange}\\\textbf{Graph}} \\
\midrule
Label            & \cmark & \cmark & \cmark & \cmark & \cmark & \cmark & \cmark & \cmark & \cmark & \xmark \\
Range            & \cmark & \cmark & \cmark & \cmark & \cmark & \xmark & \xmark & \xmark & \xmark & \cmark \\
Subset           & \cmark & \cmark & \cmark & \cmark & \cmark & \cmark & \cmark & \xmark & \cmark & \xmark \\
Boolean          & \cmark & \cmark & \cmark & \cmark & \cmark & \xmark & \xmark & \xmark & \xmark & \xmark \\
\bottomrule
\end{tabular}
\end{table*}

\footnotetext{The original filter distance used in RWalks only supports binary match/non-match filters (e.g., label and subset filters). In our experiments, we implement RWalks with our proposed filter distance, which generalizes to a much broader range of filter types and yields uniformly improved performance.}

\myparagraph{ACORN}
\cite{acorn} proposes two modified variants, ACORN-1 and ACORN-$\gamma$.
We focus on ACORN-$\gamma$ in our experiments because it achieves better search
quality. ACORN-$\gamma$ constructs a denser graph and applies two-hop expansion
during search. Suppose the minimum filter selectivity is $\gamma$ and $M$ is the
degree parameter used in standard HNSW. ACORN-$\gamma$ uses two-hop expansion to
ensure that each node has access to approximately $M \cdot \gamma$ nearby points.
After filtering out the points that do not satisfy the query filter, the
remaining neighbors should approximate those that would be obtained by building
a standard HNSW graph solely on the points satisfying the filter.

For the SIFT and LAION datasets used in the ACORN paper, we adopt the parameter
settings suggested by the authors. For the other datasets, we select parameters
from the set
$\{M = 32, M_{\beta} = 32\}$,
$\{M = 32, M_{\beta} = 64\}$,
and
$\{M = 64, M_{\beta} = 64\}$.
For $\gamma$, we set it to the minimum of the 5\% most selective queries and $30$.

\myparagraph{NaviX}
\cite{sehgal2025navix} proposes an improved search heuristic on top of ACORN.
It constructs a standard HNSW graph, and at query time adaptively decides
whether to expand one-hop or two-hop neighbors based on local selectivity
estimates. We use the authors' FAISS-Navix implementation with the
\emph{adaptive-local} heuristic enabled. The parameter configurations we test
are $\{M = 32, \texttt{efc} = 200\}$ and $\{M = 64, \texttt{efc} = 200\}$, and
we choose the better-performing configuration for each dataset.

\myparagraph{FilteredVamana}
\cite{gollapudi2023filtered} proposes an extended version of the Vamana algorithm
tailored to subset-filtered search. The original experiments focus on the
single-filter case, which naturally generalizes to multiple OR-filters. In our
evaluation, we also test the algorithm on multiple AND-filters (our subset
setting). The key idea is that when inserting a point, the algorithm traverses
and connects only to those existing points that share at least one common
attribute with the inserted point. During their ``FilteredRobustPruning'', if an edge $(p, p')$ is
pruned in favor of $(p, p^*)$, the replacement point $p^*$ must cover all the
attributes shared between $p$ and $p'$. At query time, the algorithm restricts
traversal to points that satisfy the query filter. We select optimal parameters
from, $\texttt{deg} \in \{64, 96, 128\}\,$ and $\,\alpha = 1.2$.

\myparagraph{StitchedVamana}
This is the second algorithm introduced in \cite{gollapudi2023filtered}. It is
primarily designed for the single-filter setting, but we also evaluate it under
multiple AND-filters (our subset case). For each distinct label in the dataset,
the algorithm builds a graph over the points containing that label using a small
degree parameter $R_{\text{small}}$. For points with multiple labels, it merges
the outgoing edges from the corresponding per-label graphs using
\emph{FilteredRobustPruning} with a larger degree $R_{\text{stitched}}$.
During query processing, traversal is restricted to points satisfying the filter.
The parameter configurations we test are
$R_{\text{stitched}} \in \{64, 96, 128\}$,
$R_{\text{small}} = R_{\text{stitched}}/2$,
and $L = 1.5 \cdot R_{\text{stitched}}$.

\myparagraph{RWalks}
\cite{ait2025rwalks} builds on a standard unfiltered index. For each point, the algorithm performs a random walk to compute an aggregated attribute vector. When a query arrives, it defines the filter distance between a point and the query as the fraction of labels in the aggregated attribute vector that match the query filter. In our experiments, we implement RWalks using our proposed filter distance, which strictly generalizes the binary match/non-match criterion used in the original paper and yields uniformly improved performance. During search, the greedy traversal is guided by a weighted combination of the filter distance and the standard vector distance, using a weight parameter $h$. The hyperparameter $h$ is reported to work best when set to $0.1$ (after normalization).

We implement the algorithmic ideas of RWalks on top of Vamana and introduce minor
modifications to improve performance (e.g., we use our generalized attribute and 
filter distance definitions instead of their binary equality-based definition). 
The degree parameters we test are
$\texttt{deg} \in \{64, 96, 128\}$. For all other parameters, we adopt the values
recommended in the original paper: $m = 5$, $d = 3$, $\tau = 0$, and $h = 0.1$.

\myparagraph{Post-Filtering}
We first retrieve the
nearest neighbors without applying any filter constraints, and then perform
filter checks on the results to obtain the first $k$ points that satisfy the
query filter. We test degree parameters
$\texttt{deg} \in \{32, 64, 96, 128\}$.

\myparagraph{Pre-Filtering}
We first apply the filter constraints to the points, and then compute their distances to obtain the top-$k$ results. No index is required for this Pre-Filtering method.

\myparagraph{NHQ}
\cite{NHQ} proposes the NHQ algorithm, which targets label-equality filters. It
defines an attribute distance between two points based solely on whether their
attributes are equal, and combines this with the standard vector distance using
a weighted average. Because this attribute distance can only distinguish
equality, NHQ can handle only equality-based label filters in our experiments.
We use the parameter settings recommended in \cite{arxiv-benchmark} for both the
ARXIV and SIFT datasets.

\myparagraph{UNG}
\cite{cai2024navigating-ung} introduces the Unified Navigating Graph (UNG),
designed specifically for subset-filter queries. The algorithm first constructs a
proximity graph for each possible label set. It then builds a
label-navigating graph over the label sets by connecting a directed edge whenever
one label set subsumes another. Given a subset query, UNG maintains a trie data
structure to identify feasible starting points, and then performs greedy search
from these entry points. We primarily adopt the recommended parameters from the
original paper: $\alpha = 1.2$, $\delta = 6$, and $\sigma = 16$. For the degree
$R$ and queue length $L$, we choose the best-performing configuration from
$(R, L) \in \{(32, 100), (48, 150), (64, 200)\}$.

\myparagraph{iRangeGraph}
\cite{xu2024irangegraph} proposes iRangeGraph for range-filter queries. It
constructs a segment tree to decompose the attribute range into intervals of
different lengths, builds a graph index for each interval, and searches only over
those interval graphs that overlap with the query range. We use the parameter
settings recommended by the authors: $m = 16$ and $\texttt{EF} = 100$.

\subsection{Parameter Settings}\label{sec:parameters}

We summarize the parameters used for each algorithm and dataset below.

\begin{itemize}
    \item Threshold-\algoname.  
    We use $\alpha = 1.2$ for all datasets.
    \begin{itemize}
        \item \textbf{SIFT-1M (label):} $\text{deg} = 96$, $\text{Thresholds} = \{10\%, 0\}$
        \item \textbf{ARXIV (label):} $\text{deg} = 64$, $\text{Thresholds} = \{100\%, 0\}$
        \item \textbf{ARXIV (range):} $\text{deg} = 96$, $\text{Thresholds} = \{100\%, 10\%\}$
        \item \textbf{LAION-1M / LAION-5M / LAION-25M:} $\text{deg} = 96$, $\text{Thresholds} = \{10\%, 1\%, 0\}$
        \item \textbf{YFCC:} $\text{deg} = 128$, $\text{Thresholds} = \{10\%, 1\%, 0\}$
        \item \textbf{MSTuring (subset, range, boolean):} $\text{deg} = 128$, $\text{Thresholds} = \{100\%, 1\%, 0\}$
    \end{itemize}

    \item Weight-\algoname.  
    We use $\alpha = 1.2$ for all datasets.
    \begin{itemize}
        \item \textbf{SIFT-1M (label):} $\text{deg} = 96$, $\text{Weight} = \{1\}$
        \item \textbf{ARXIV (label):} $\text{deg} = 64$, $\text{Weight} = \{0,10\}$
        \item \textbf{ARXIV (range):} $\text{deg} = 96$, $\text{Weight} =\{0,2\}$
        \item \textbf{LAION-1M / LAION-5M / LAION-25M (subset):} $\text{deg} = 96$, $\text{Weight} =\{0,1\}$
        \item \textbf{YFCC (subset):} $\text{deg} = 128$, $\text{Weight} =\{0,1\}$
        \item \textbf{MSTuring (subset):} $\text{deg} = 128$, $\text{Weight} =\{0,1,5\}$
        \item \textbf{MSTuring (range):} $\text{deg} = 128$, $\text{Weight} =\{0,10,1000\}$
        \item \textbf{MSTuring (boolean):} $\text{deg} = 128$, $\text{Weight} =\{0,1,10\}$
    \end{itemize}
    \item ACORN.  
    \begin{itemize}
        \item \textbf{SIFT-1M (label):} $M = 32$, $\gamma = 12$, $M_{\beta} = 64$
        \item \textbf{ARXIV (label):} $M = 32$, $\gamma = 25$, $M_{\beta} = 32$
        \item \textbf{ARXIV (range):} $M = 32$, $\gamma = 10$, $M_{\beta} = 32$
        \item \textbf{LAION-1M / LAION-5M / LAION-25M (subset):} $M = 32$, $\gamma = 30$, $M_{\beta} = 32$
        \item \textbf{YFCC (subset):} $M = 64$, $\gamma = 30$, $M_{\beta} = 64$
        \item \textbf{MSTuring (subset):} $M = 64$, $\gamma = 30$, $M_{\beta} = 64$
        \item \textbf{MSTuring (range):} $M = 64$, $\gamma = 30$, $M_{\beta} = 64$
        \item \textbf{MSTuring (boolean):} $M = 32$, $\gamma = 30$, $M_{\beta} = 64$
    \end{itemize}
    \item NaviX.  
    \begin{itemize}
        \item \textbf{SIFT-1M (label):} $M = 32$, $efc = 200$
        \item \textbf{ARXIV (label):} $M = 64$, $efc = 200$
        \item \textbf{ARXIV (range):} $M = 64$, $efc = 200$
        \item \textbf{LAION-1M / LAION-5M / LAION-25M (subset):} $M = 32$, $efc = 200$
        \item \textbf{YFCC (subset):} $M = 64$, $efc = 200$
        \item \textbf{MSTuring (subset):} $M = 64$, $efc = 200$
        \item \textbf{MSTuring (range):} $M = 64$, $efc = 200$
        \item \textbf{MSTuring (boolean):} $M = 64$, $efc = 200$
    \end{itemize}
    \item RWalks. $\alpha=1.2$, $m=5$, $d=3$, $\tau=0$, $h=0.1$ 
    \begin{itemize}
        \item \textbf{SIFT-1M (label):} $\text{deg} = 64$
        \item \textbf{ARXIV (label):} $\text{deg} = 64$
        \item \textbf{ARXIV (range):} $\text{deg} = 64$
        \item \textbf{LAION-1M (subset) :} $\text{deg} = 64$
        \item \textbf{LAION-5M / LAION-25M (subset):} $\text{deg} = 96$
        \item \textbf{YFCC (subset):} $\text{deg} = 96$
        \item \textbf{MSTuring (subset):} $\text{deg} = 96$
        \item \textbf{MSTuring (range):} $\text{deg} = 128$
        \item \textbf{MSTuring (boolean):} $\text{deg} = 96$
    \end{itemize}
    \item Post-Filtering. $\alpha=1.2$
    \begin{itemize}
        \item \textbf{SIFT-1M (label):} $\text{deg} = 32$
        \item \textbf{ARXIV (label):} $\text{deg} = 96$
        \item \textbf{ARXIV (range):} $\text{deg} = 32$
        \item \textbf{LAION-1M (subset) :} $\text{deg} = 32$
        \item \textbf{LAION-5M (subset):} $\text{deg} = 64$
        \item \textbf{LAION-25M (subset):} $\text{deg} = 128$
        \item \textbf{YFCC (subset):} $\text{deg} = 64$
        \item \textbf{MSTuring (subset):} $\text{deg} = 96$
        \item \textbf{MSTuring (range):} $\text{deg} = 64$
        \item \textbf{MSTuring (boolean):} $\text{deg} = 64$
    \end{itemize}
    \item FilteredVamana. $\alpha=1.2$
    \begin{itemize}
        \item \textbf{SIFT-1M (label):} $\text{deg} = 64$
        \item \textbf{ARXIV (label):} $\text{deg} = 96$
        \item \textbf{LAION-1M (subset) :} $\text{deg} = 96$
        \item \textbf{LAION-5M (subset):} $\text{deg} = 96$
        \item \textbf{LAION-25M (subset):} $\text{deg} = 128$
        \item \textbf{YFCC (subset):} $\text{deg} = 128$
        \item \textbf{MSTuring (subset):} $\text{deg} = 128$
    \end{itemize}
    \item StitchedVamana. $\alpha=1.2$
    \begin{itemize}
        \item \textbf{SIFT-1M (label):} $\text{deg} = 64$
        \item \textbf{ARXIV (label):} $\text{deg} = 96$
        \item \textbf{LAION-1M (subset) :} $\text{deg} = 96$
        \item \textbf{LAION-5M (subset):} $\text{deg} = 96$
        \item \textbf{LAION-25M (subset):} $\text{deg} = 96$
        \item \textbf{YFCC (subset):} $\text{deg} = 64$
        \item \textbf{MSTuring (subset):} $\text{deg} = 128$
    \end{itemize}
    \item NHQ
    \begin{itemize}
        \item \textbf{SIFT-1M (label):} $K = 80, L = 60, S = 10, R = 200, \text{RANGE} = 60, PL = 300, B = 0.6, \text{weight\_search} = 1000000$
        \item \textbf{ARXIV (label):} $K = 80, L = 60, S = 10, R = 200, \text{RANGE} = 60, PL = 300, B = 0.6, \text{weight\_search} = 1000000$
    \end{itemize}
    \item UNG. $\alpha=1.2$, $\delta = 6$, $\sigma = 16$
    \begin{itemize}
        \item \textbf{SIFT-1M (label):} $\text{deg} = 32$, $\text{L} = 100$
        \item \textbf{ARXIV (label):} $\text{deg} = 48$, $\text{L} = 150$
        \item \textbf{LAION-1M (subset):} $\text{deg} = 32$, $\text{L} = 100$
        \item \textbf{LAION-5M (subset):} $\text{deg} = 32$, $\text{L} = 100$
        \item \textbf{LAION-25M (subset):} $\text{deg} = 32$, $\text{L} = 100$
        \item \textbf{YFCC (subset):} $\text{deg} = 32$, $\text{L} = 100$
    \end{itemize}
    \item iRangeGraph.
    \begin{itemize}
        \item \textbf{ARXIV (range):} $M = 16, ef = 100$
        \item \textbf{MSTuring (range):} $M = 16, ef = 100$
    \end{itemize}
\end{itemize}

\subsection{Indexing Time}

We report the indexing time in Table~\ref{tab:indexing-time} for all algorithms and datasets using the parameters specified in Section \ref{sec:parameters}. As described in Section \ref{sec:exp-alg}, our parameter-selection guideline is to maximize QPS at 0.95 recall, rather than to optimize indexing time or index size.

\begin{table*}[]
\caption{Indexing time for all algorithms across different datasets. We show the indexing time for Threshold-\algoname; the indexing time for Weight-\algoname\ is similar. Please refer to Appendix~\ref{sec:parameters} for the parameters used to build the index.}
\label{tab:indexing-time}
\begin{tabular}{c|cccccccccc}
\toprule
& \algoname{}
& ACORN  
& NaviX 
& \makecell{Filtered\\Vamana} 
& \makecell{Stitched\\Vamana} 
& \makecell{Post\\Filtering} 
& \makecell{RWalks} 
& UNG 
& NHQ 
& \makecell{iRange\\Graph} \\
\midrule
SIFT-1M-label       & 100s  & 222s & 54s  & 24s            & 26s            & 11s            & 122s & 15s & 28s  & N/A \\     
ARXIV-2M-label       & 3174s  & 16016s & 2378s  & 3011s            & 5599s            & 3409s            & 19685s & 3779s & 515s & N/A \\
ARXIV-2M-range       & 5978s  & 5211s & 2081s  & N/A            & N/A            & 750s            & 19642s & N/A & N/A & 7424s \\
LAION-1M-subset     & 261s  & 1148s & 145s & 211s           & 1249s          & 31s            & 365s & 13s & N/A & N/A \\  
LAION-5M-subset     & 1537s & 6960s & 872s & 1232s          & 3683s          & 328s           & 3293s & 89s & N/A & N/A \\    
LAION-25M-subset    & 8687s & 45662s & 4981s & 9401s          & 36179s         & 3209s          & 17733s & 569s & N/A & N/A \\  
YFCC-10M-subset    & 7898s & 22572s & 1220s & 4515s          & 42812s         & 300s          & 3355s & 8453s & N/A & N/A\\   
MSTuring-10M-subset & 3760s & 21742s & 1276s & 1257s          & 33376s         & 500s           & 4140s  & $>$15h & N/A & N/A\\    
MSTuring-10M-range & 3348s & 21322s & 1281s & N/A          & N/A         & 311s           & 7228s  & N/A & N/A & 6596s\\
MSTuring-10M-bool & 3760s & 21185s  & 1445s & N/A          & N/A         & 310s           & 4140s  & N/A & N/A & N/A \\
\bottomrule
\end{tabular}

\end{table*}

\subsection{Distance Computation v.s. Recall}\label{sec:exp-dc}

\begin{figure*}[t]
    \centering
    \begin{subfigure}[t]{0.43\textwidth}
        \centering
        \includegraphics[width=\linewidth]{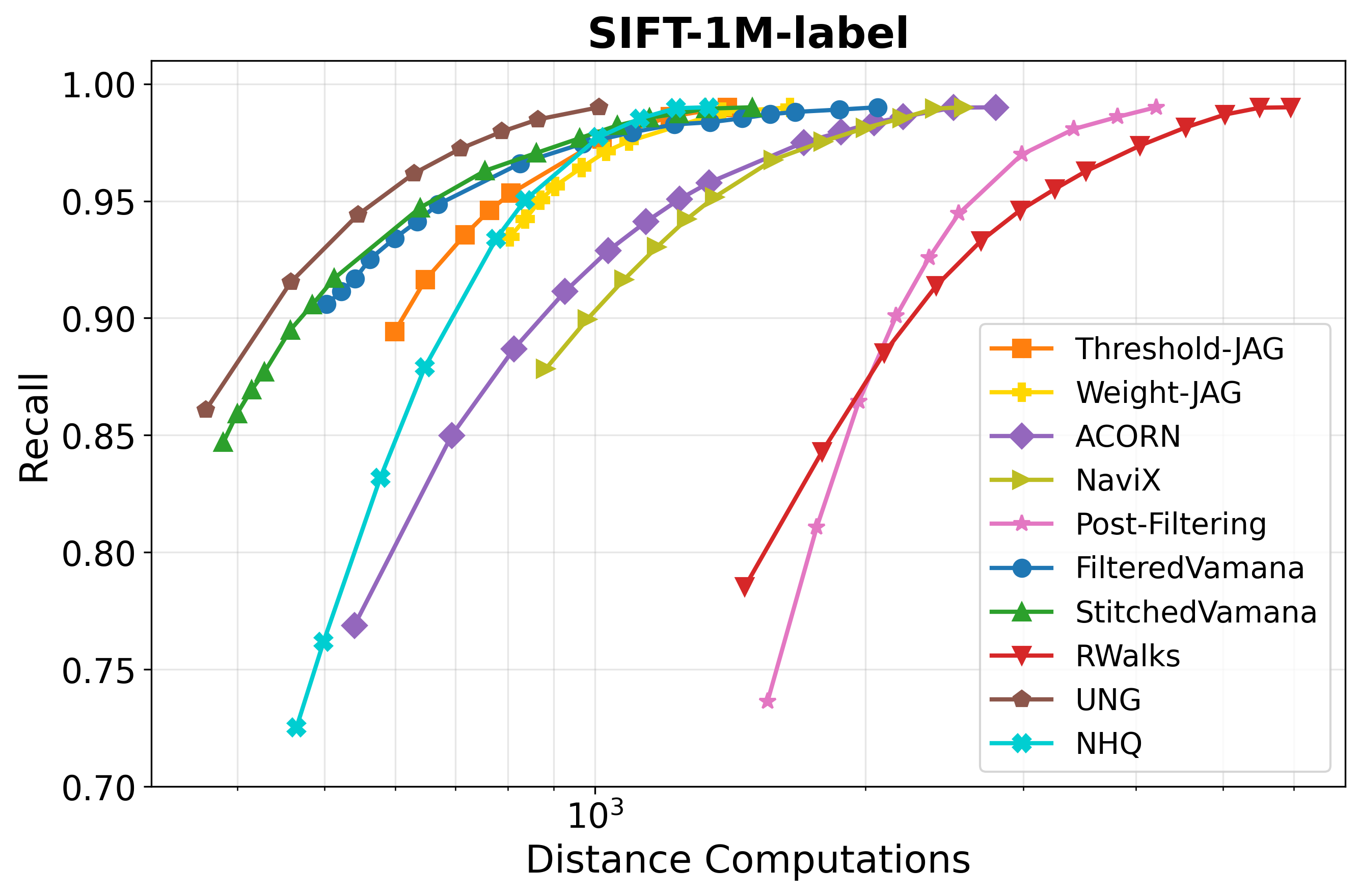}
        \label{fig:sift-label-dc}
    \end{subfigure}
    \hfill
    \begin{subfigure}[t]{0.43\textwidth}
        \centering
        \includegraphics[width=\linewidth]{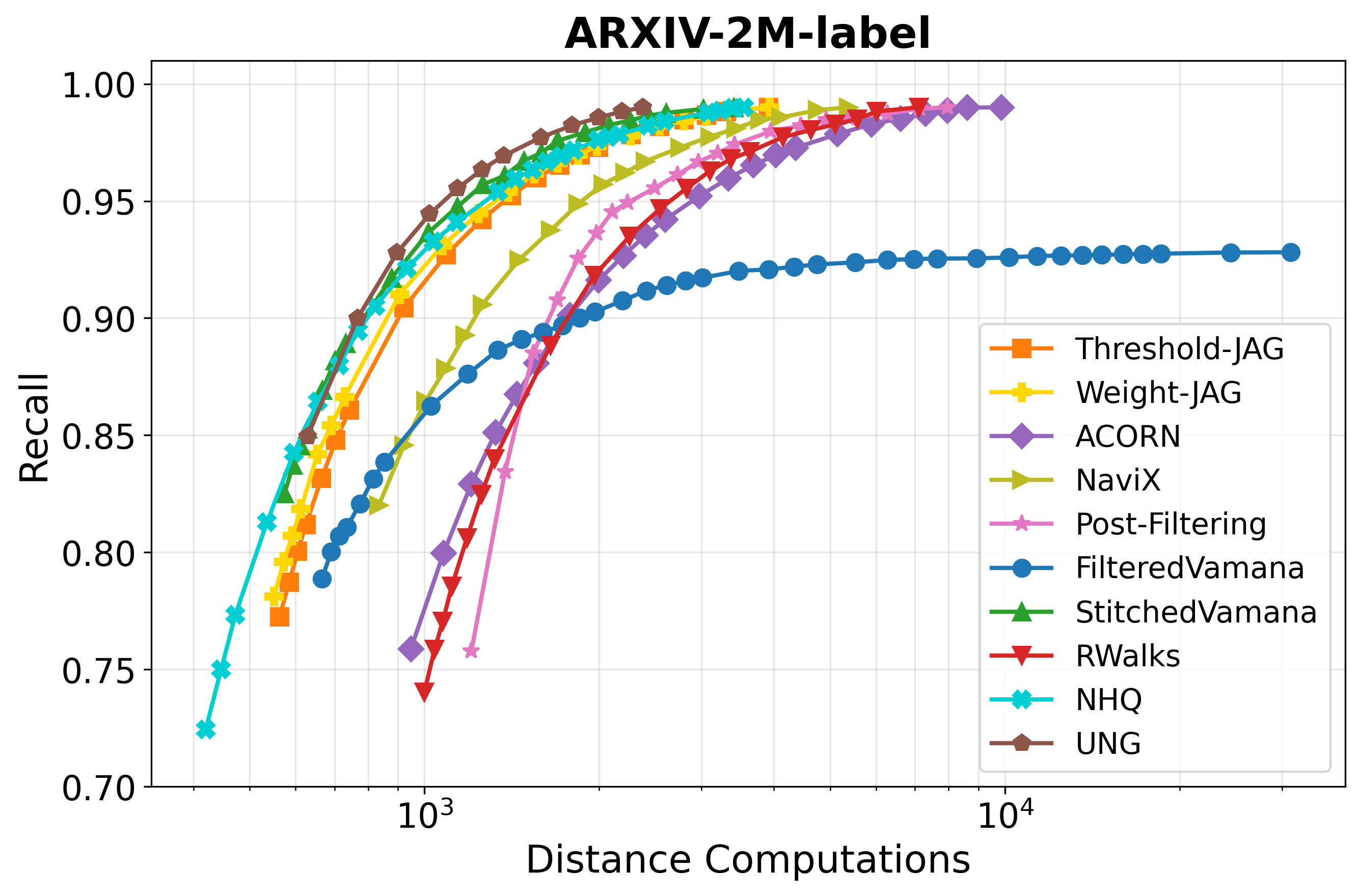}
        \label{fig:arxiv-label-dc}
    \end{subfigure}
    \caption{Distance computation vs. recall plot for label filters on the SIFT and ARXIV datasets. Note that NHQ is designed specifically for Label filter. FilteredVamana, StitchedVamana, and UNG are designed specifically for Label and Subset filters.}
    \label{fig:exp-label-dc}
\end{figure*}

\begin{figure*}[t]
    \centering
    \begin{subfigure}[t]{0.43\textwidth}
        \centering
        \includegraphics[width=\linewidth]{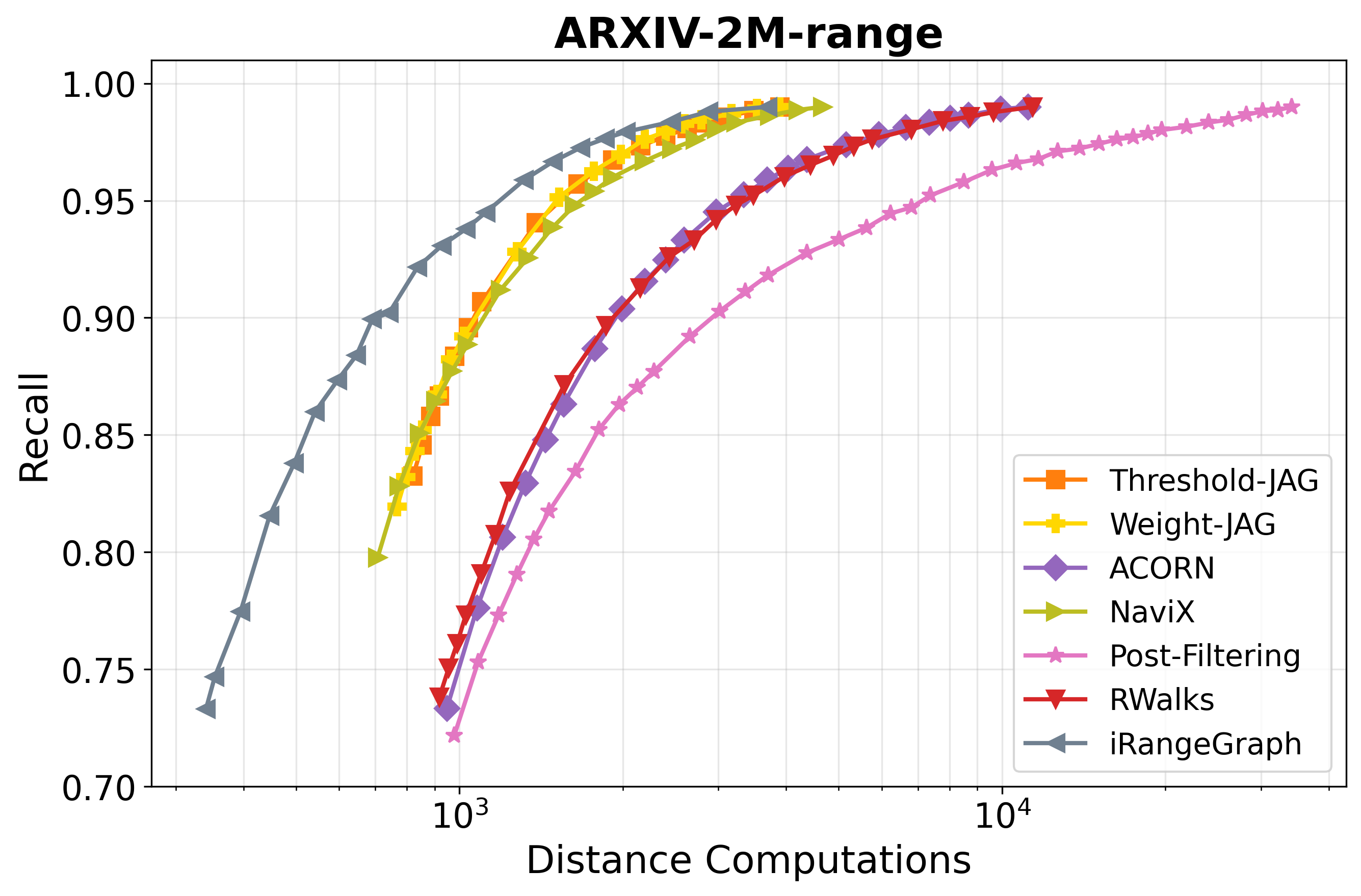}
        \label{fig:arxiv-range-dc}
    \end{subfigure}
    \hfill
    \begin{subfigure}[t]{0.43\textwidth}
        \centering
        \includegraphics[width=\linewidth]{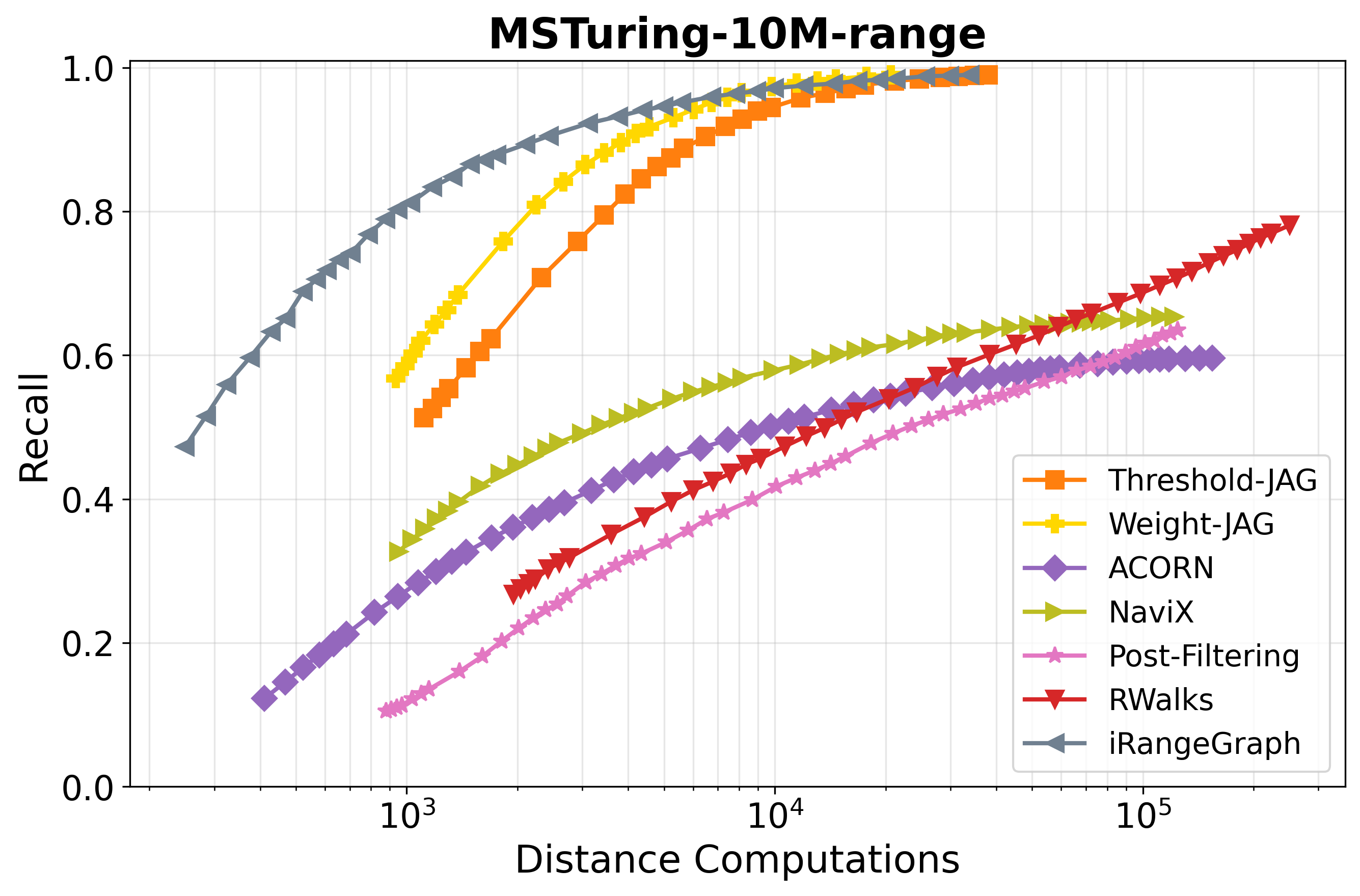}
        \label{fig:msturing-range-dc}
    \end{subfigure}
    \caption{Distance computation vs. recall plot for range filters on the ARXIV and MSTuring datasets. Note that iRangeGraph is designed specifically for Range filter}
    \label{fig:exp-range-dc}
\end{figure*}

\begin{figure*}[t]
    \centering
    \begin{subfigure}[t]{0.43\textwidth}
        \centering
        \includegraphics[width=\linewidth]{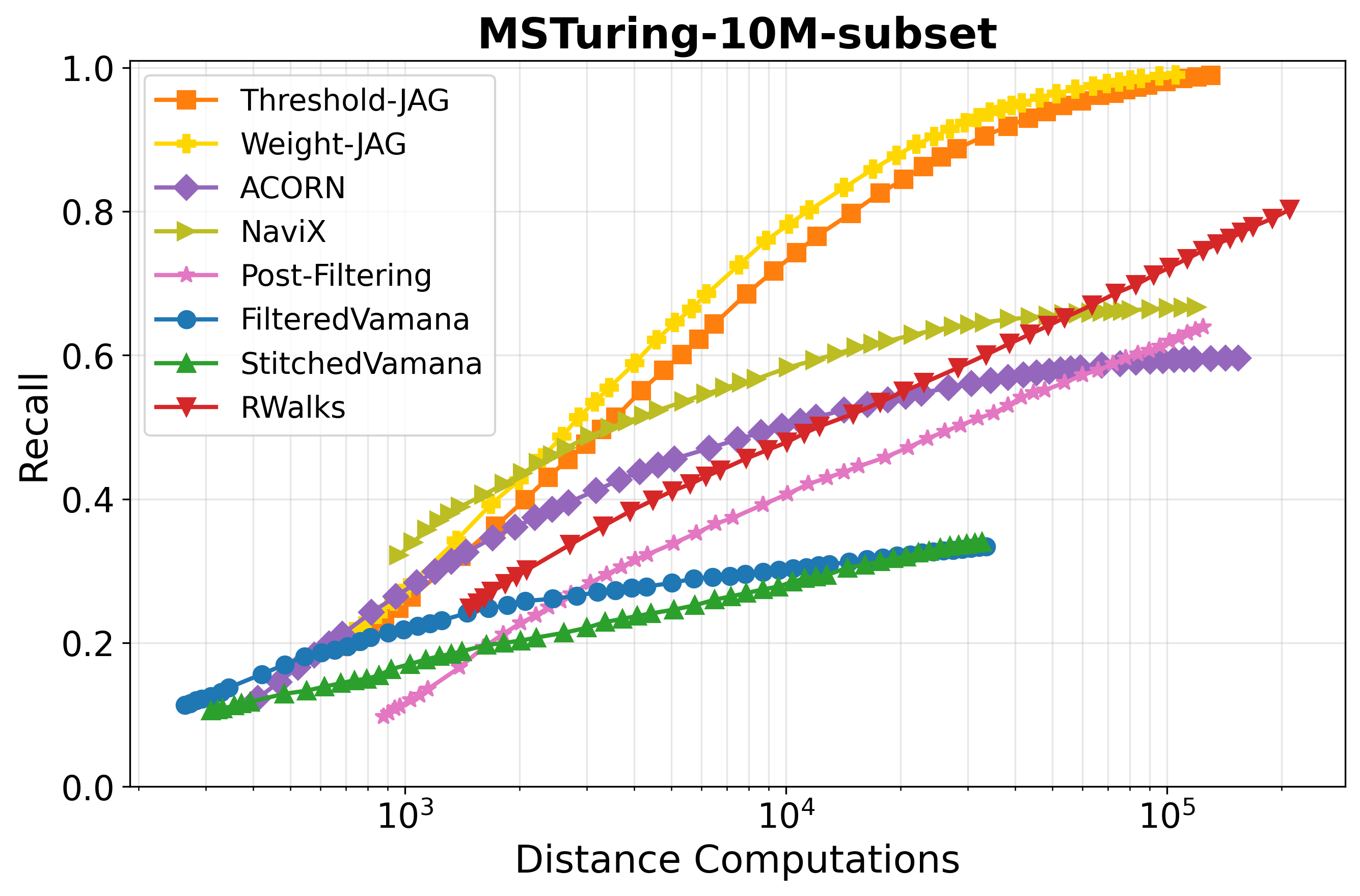}
        \label{fig:msturing-subset-dc}
    \end{subfigure}
    \hfill
    \begin{subfigure}[t]{0.43\textwidth}
        \centering
        \includegraphics[width=\linewidth]{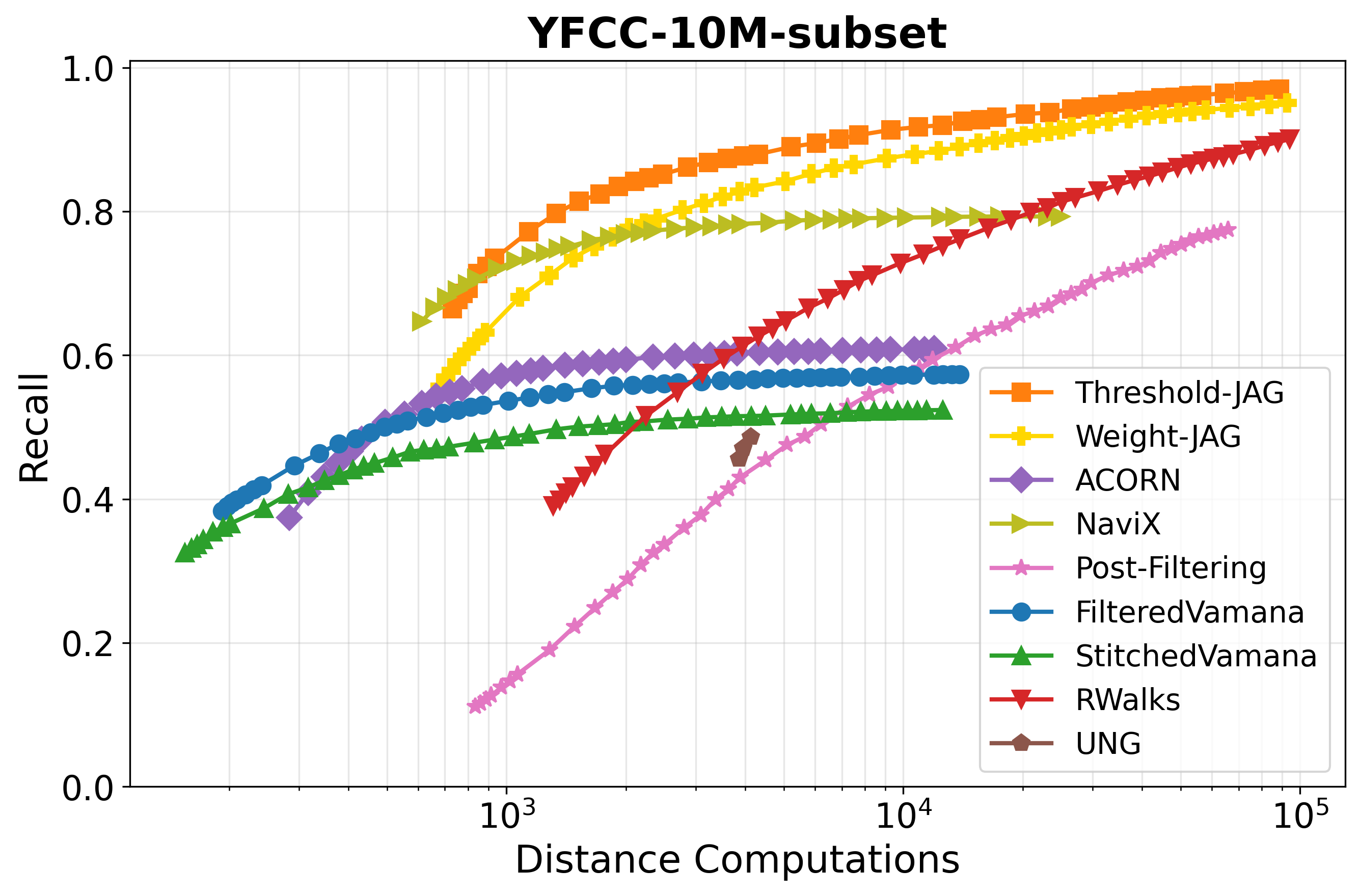}
        \label{fig:yfcc-subset-dc}
    \end{subfigure}

    \vspace{1em}
    
    \begin{subfigure}[t]{0.43\textwidth}
        \centering
        \includegraphics[width=\linewidth]{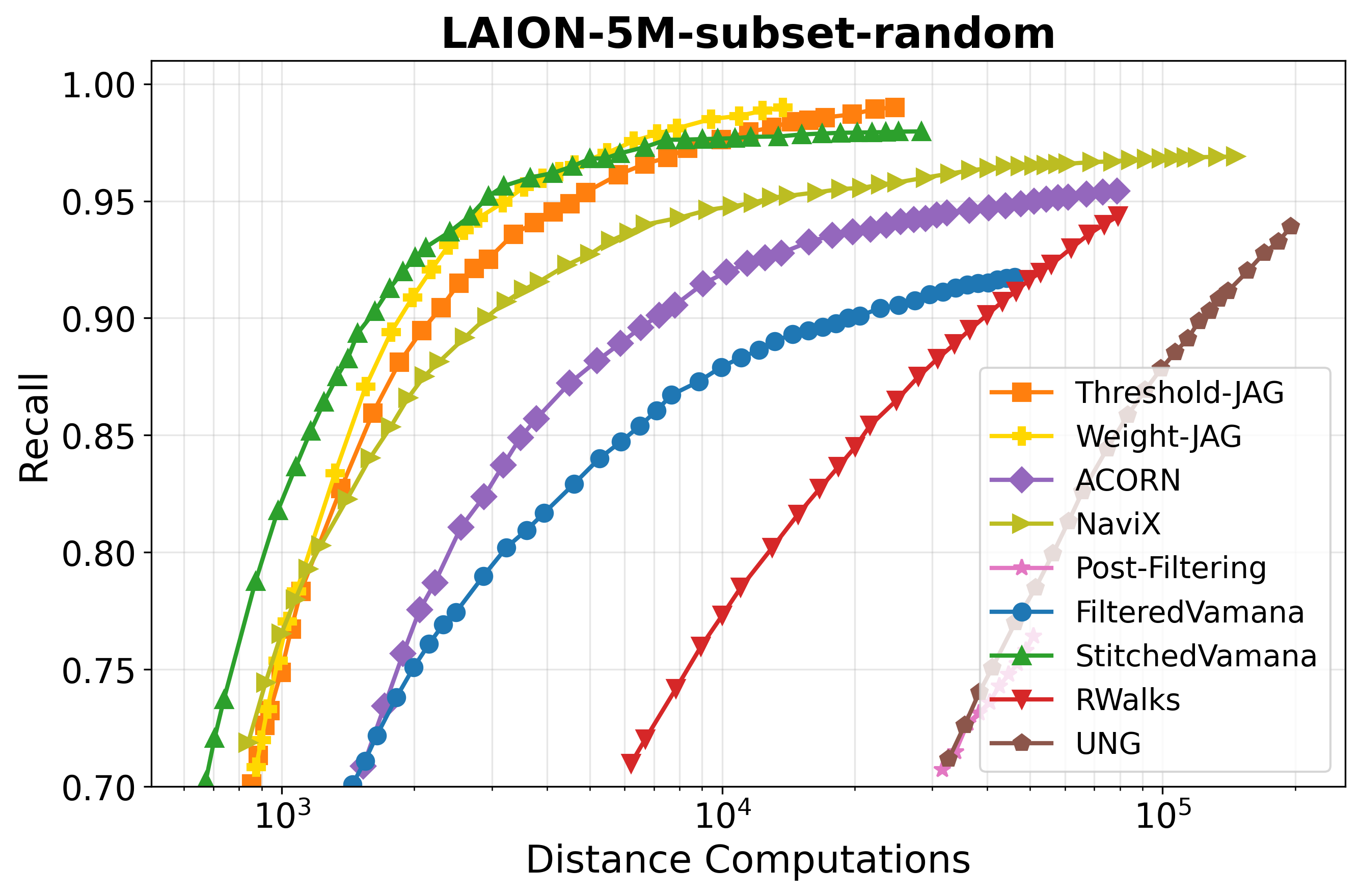}
        \label{fig:laion5m-subset-dc}
    \end{subfigure}
    \hfill
    \begin{subfigure}[t]{0.43\textwidth}
        \centering
        \includegraphics[width=\linewidth]{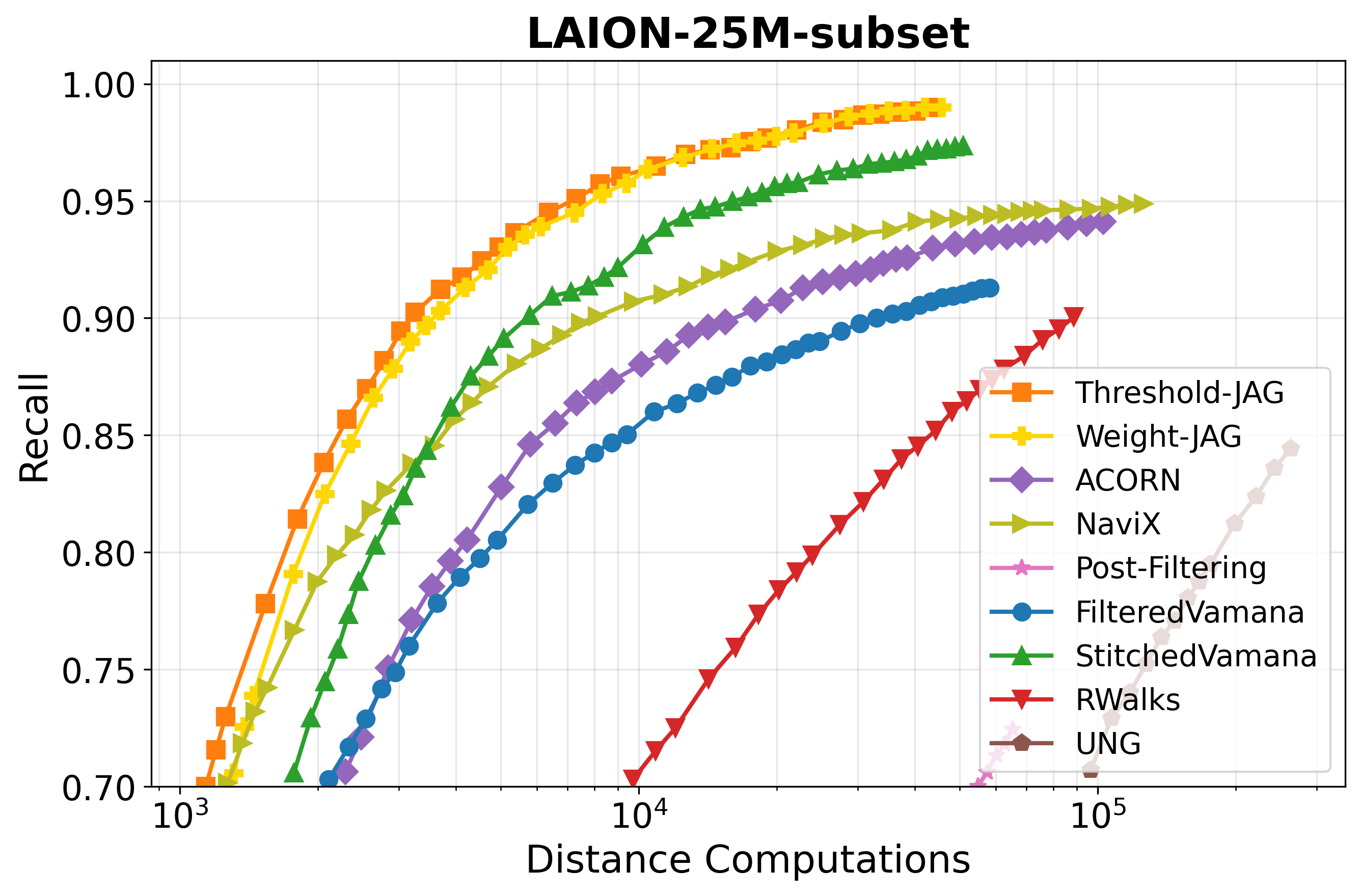}
        \label{fig:laion25m-subset-dc}
    \end{subfigure}
    \caption{Distance computation vs. recall plots for subset filters on the MSTuring-10M, LAION-5M, LAION-25M, and YFCC dataset. Note that FilteredVamana, StitchedVamana, and UNG are designed specifically for label and subset filters.}
    \label{fig:exp-subset-dc}
\end{figure*}

\begin{figure*}[h]
    \centering
    \includegraphics[width=0.43\textwidth]{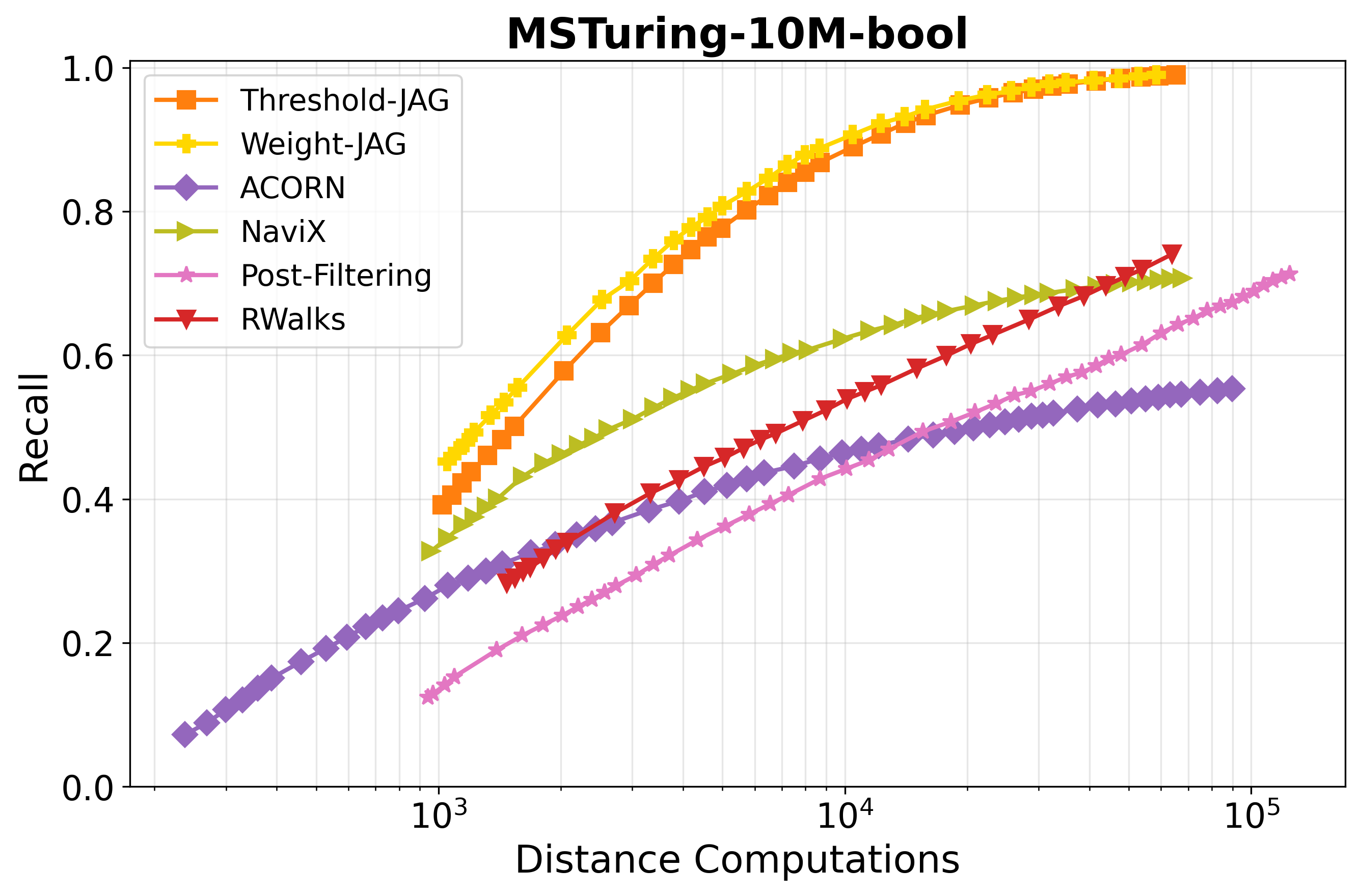}
    \caption{Distance computation vs. recall for boolean filters on the MSTuring dataset.}
    \label{fig:exp-bool-dc}
\end{figure*}

We report recall@10 versus distance computation in Figure~\ref{fig:exp-label-dc},~\ref{fig:exp-range-dc},~\ref{fig:exp-subset-dc},~\ref{fig:exp-bool-dc}.

